\documentstyle[12pt,epsfig,twoside]{article}
\newcommand{\bei}{\begin{itemize}}
\newcommand{\eei}{\end{itemize}}
\newcommand{\beq}{\begin{equation}}
\newcommand{\eeq}{\end{equation}}
\newcommand{\beqn}{\begin{eqnarray}}
\newcommand{\eeqn}{\end{eqnarray}}
\newcommand{\beqns}{\begin{eqnarray*}}
\newcommand{\eeqns}{\end{eqnarray*}}
\newcommand{\vs}{\\[0.3cm]\indent}

\newcommand{\intl}{\int\limits}

\newcommand{\mc}{\multicolumn}

\def\NP{{\it Nucl. Phys.}}
\def\PL{{\it Phys. Lett.}}

\def\PRL{{\it Phys. Rev. Lett.}}

\def\ea{{\it et al.}}
\def\pc{$\%$}

\def\sf{spectral function}
\def\sfs{spectral functions}

\def\ee{$e^+e^-$}

\def\aqedZ{$\alpha(M_{Z}^2)$}

\def\daqedZ{$\Delta\alpha(M_{Z}^2)$}

\def\daqedhZ{$\Delta\alpha_{\rm had}(M_{Z}^2)$}

\def\amuhadLO{$a_\mu^{\rm had,LO}$}
\def\tauto{$\tau^{-\!}\rightarrow\,$}
\def\nut{$\,\nu_\tau$}
\def\piz{$ \pi^0 $}
\def\pipiz{$ \pi^-\pi^0 $}
\def\pitpiz{$ \pi^-3\pi^0 $}
\def\tpipiz{$ 2\pi^-\pi^+\pi^0 $}
\def\rar{\rightarrow}

\def\Ks{$K^0_{\mathrm S}$}
\def\Kl{$K^0_{\mathrm L}$}

\def\ie{{\it i.e.}} 
 
\def\via{via} 

\def\cf{{\em cf.}}
\def\rs{\raisebox{1.5ex}[-1.5ex]}

\begin{document}

\begin{titlepage}
\setcounter{page}{1}

\begin{flushright} 
{\bf LAL 02-81}\\
hep-ph/0208177.v3 \\
\today
\end{flushright} 

\vspace{-0.2cm}

\begin{center} 
\begin{Large}
{\bf Confronting Spectral Functions from \boldmath\ee\ Annihilation 
and \boldmath$\tau$ Decays:
Consequences for the Muon Magnetic Moment} \\
\end{Large}
\vspace{1.0cm}
\begin{large}
M.~Davier$^{\,\mathrm a}$,
S.~Eidelman$^{\,\mathrm b}$,
A.~H\"ocker$^{\,\mathrm a}$
and Z.~Zhang$^{\,\mathrm a,}$\footnote
{
	E-mail: 
	davier@lal.in2p3.fr,
	simon.eidelman@cern.ch,
	hoecker@lal.in2p3.fr,
	zhangzq@lal.in2p3.fr
} \\
\end{large}
\vspace{0.5cm}
{\small \em $^{\mathrm a}$Laboratoire de l'Acc\'el\'erateur Lin\'eaire,\\
IN2P3-CNRS et Universit\'e de Paris-Sud, F-91898 Orsay, France}\\
\vspace{0.1cm}
{\small \em $^{\mathrm b}$Budker Institute
			  of Nuclear Physics, Novosibirsk, 630090, Russia }\\
\vspace{1.0cm}

{\small{\bf Abstract}}
\end{center}
{\small
\vspace{-0.2cm}
Vacuum polarization integrals involve the vector spectral functions which
can be experimentally determined from two sources:
({\it i}) \ee\ annihilation cross sections and ({\it ii}) hadronic $\tau$ 
decays. Recently results with comparable precision have become available 
from CMD-2 on one side, and ALEPH, CLEO and OPAL on the other. The comparison
of the respective spectral functions involves a correction from
isospin-breaking effects, which is evaluated. After the correction
it is found that the dominant $\pi\pi$ spectral functions do not 
agree within experimental and theoretical uncertainties. Some 
disagreement is also found for the $4\pi$ spectral functions. 
The consequences of these discrepancies for vacuum polarization
calculations are presented, with the emphasis on
the muon anomalous magnetic moment. The work includes a complete 
re-evaluation of all exclusive cross sections, taking into account
the most recent data that became available in particular from the 
Novosibirsk experiments and applying corrections for the missing
radiative corrections. The values found for the lowest-order
hadronic vacuum polarization contributions are
\beqns
a_\mu^{\rm had,LO} = \left\{
\begin{array}{ll}
     	 (684.7\pm6.0_{\rm exp}\pm3.6_{\rm rad})~10^{-10}
     	& ~~~[e^+e^-{\rm -based}]~,\\[0.1cm]
	 (709.0\pm5.1_{\rm exp}\pm1.2_{\rm rad}
				\pm2.8_{\rm SU(2)})~10^{-10}
     & ~~~[\tau {\rm -based}]~, \\
\end{array}
\right.
\eeqns
where the errors have been separated according to their sources: 
experimental, missing radiative corrections in \ee\ data, and 
isospin breaking.  The Standard Model predictions for the muon 
magnetic anomaly read
\beqns
a_\mu = \left\{
\begin{array}{ll}
	(11\,659\,169.3\pm7.0_{\rm had}
		\pm3.5_{\rm LBL}\pm0.4_{\rm QED+EW})~10^{-10}
     & ~~~[e^+e^-{\rm -based}]~, \\[0.1cm]
	(11\,659\,193.6\pm5.9_{\rm had}
		\pm3.5_{\rm LBL}\pm0.4_{\rm QED+EW})~10^{-10}
     & ~~~[\tau {\rm -based}]~, \\
\end{array}
\right.
\eeqns
where the errors account for the hadronic, light-by-light scattering
and electroweak contributions. We observe deviations with the recent 
BNL measurement
at the 3.0 (\ee) and 0.9 ($\tau$) $\sigma$ level, when adding 
experimental and theoretical errors in quadrature.
\noindent
}
\vspace{25mm}

\end{titlepage}

\tableofcontents
\vfill
\pagebreak

\setcounter{page}{1}
%
%
\section{Introduction}
\label{sec_introduction}

Hadronic vacuum polarization in the photon propagator plays an important 
role in the precision tests of the Standard Model. This is the case for the
evaluation of the electromagnetic coupling at the $Z$ mass scale,
$\alpha (M_Z^2)$, which receives a contribution 
$ \Delta\alpha_{\rm had}(M_{Z}^2)$ of the order of $2.8~10^{-2}$ 
that must be known to an accuracy of better than 1\% so that it does not
limit the accuracy on the indirect determination of the Higgs boson mass
from the measurement of $\sin ^2 \theta_W$. Another example is provided
by the anomalous magnetic moment $a_\mu=(g_\mu -2)/2$ of the muon where the
hadronic vacuum polarization component is the leading contributor to the
uncertainty of the theoretical prediction.
\vs
Starting from Refs.~\cite{cabibbo,bouchiat} there is a long history of
calculating the contributions from hadronic vacuum polarization 
in these processes. As they cannot be obtained 
from first principles because of the low energy scale
involved, the computation relies on analyticity and unitarity so that the
relevant integrals can be expressed in terms of an experimentally determined
spectral function which is proportional to the cross section for \ee\
annihilation into hadrons. The accuracy of the calculations has therefore
followed the progress in the quality of the corresponding data~\cite{eidelman}.
Because the latter was not always suitable, it was deemed necessary to resort 
to other sources of information. One such possibility was the 
use~\cite{adh} of the vector spectral functions derived from the study 
of hadronic $\tau$ decays~\cite{aleph_vsf} for the energy range less 
than 1.8~GeV. Another one occurred when it was realized in the study of 
$\tau$ decays~\cite{aleph_asf} that perturbative QCD could be applied to 
energy scales as low as 1-2~GeV, thus offering a way to replace poor \ee\ 
data in some energy regions by a reliable and precise theoretical 
prescription~\cite{dh97,steinhauser,martin}. Finally, without any further 
theoretical assumption, it was proposed to use QCD sum rules~\cite{groote,dh98}
in order to improve the evaluation in energy regions dominated by
resonances where one has to rely on experimental data. 
Using these improvements the lowest-order hadronic contribution 
to $a_\mu$ was found to be~\cite{dh98}
\beq
    a_\mu^{\rm had,LO} \:=\: (692.4 \pm 6.2)~10^{-10}~.
\label{dh98}
\eeq
The complete theoretical prediction includes in addition QED, 
weak and higher order hadronic contributions.
\vs
The anomalous magnetic moment of the muon is experimentally
known to very high accuracy. Combined with the older less precise results
from CERN~\cite{bailey}, the measurements from the E821 experiment at 
BNL~\cite{carey,brown,bnl},
including the most recent result~\cite{bnl_2002},
 yield 
\beq
    a_\mu^{\rm exp} \:=\: (11\,659\,203 \pm 8)~10^{-10}~,
\label{bnl}
\eeq
and are aiming at an ultimate precision of $4~10^{-10}$ in the future. 
The previous experimental result~\cite{bnl} was found to 
deviate from the theoretical prediction by 2.6~$\sigma$, 
but a large part of the discrepancy was actually
originating from a sign mistake in the calculation of the small
contribution from the so-called light-by-light (LBL) scattering 
diagrams~\cite{kino_light,bij_light}. The new calculations of the
LBL contribution~\cite{knecht_light,kino_light_cor,bij_light_cor} have
reduced the discrepancy to a nonsignificant 1.6~$\sigma$ level. At any rate
it is clear that the presently achieved experimental accuracy already 
calls for a more precise evaluation of \amuhadLO.
\vs
In this paper we critically review the available experimental input to
vacuum polarization integrals. Such a re-evaluation is necessary 
because
\bei
\item new results have been obtained at Novosibirsk with the CMD-2 
detector in the region dominated by the $\rho$ resonance~\cite{cmd2} 
with a much higher precision 
than before, and more accurate R measurements have been 
performed in Beijing with the BES detector in the 2-5 GeV energy 
range~\cite{bes}.
\item new preliminary results are available from the final analysis of 
$\tau$ decays with ALEPH using the full statistics accumulated at 
LEP1~\cite{aleph_new}; also the information from the spectral functions
measured by CLEO~\cite{cleo_2pi,cleo_4pi} and OPAL~\cite{opal} was not 
used previously and can be incorporated in the analysis. 
\item new results on the evaluation of isospin breaking have been 
produced~\cite{czyz,ecker1,ecker2}, thus providing a better understanding of
this critical area when relating vector $\tau$ and isovector \ee\ 
spectral functions.
\eei

Since we are mostly dealing with the low energy region, common to both
\ee\ and $\tau$ data, and because of the current interest in the muon
magnetic moment prompted by the new experimental result,
the emphasis in this paper is on \amuhadLO\ rather
than \daqedhZ. It is true that the presently achieved accuracy on
\daqedhZ\ is meeting the goals for the LEP/SLD/FNAL global electroweak
fit. However the situation will change in the long run when very
precise determinations of $\sin ^2 \theta_{\rm W}$, as could be available 
from the beam polarization asymmetry at the future Linear Collider, 
necessitate a significant increase of the accuracy on \daqedhZ~\cite{linear}.
\vs
{\bf Disclaimer:} 'theoretical' predictions using vacuum polarization 
integrals are based on experimental data as input. The data 
incorporated in this analysis are used as quoted by their authors. 
In particular, no attempt has been made to re-evaluate systematic 
uncertainties even if their size was deemed to be questionable in 
some cases. However, whenever significant incompatibilities 
between experiments occur, we apply an appropriate rescaling of the
combined error. The analysis thus heavily relies on the quality
of the work performed in the experiments.

%
%
%
%
\section{Muon Magnetic Anomaly}
\label{anomaly}

It is convenient to separate the Standard Model prediction for the
anomalous magnetic moment of the muon
into its different contributions,
\beq
    a_\mu^{\mathrm SM} \:=\: a_\mu^{\mathrm QED} + a_\mu^{\mathrm had} +
                             a_\mu^{\mathrm weak}~,
\eeq
with
\beq
 a_\mu^{\mathrm had} \:=\: a_\mu^{\mathrm had,LO} + a_\mu^{\mathrm had,HO}
           + a_\mu^{\mathrm had,LBL}~,
\eeq
where $a_\mu^{\mathrm QED}=(11\,658\,470.6\pm0.3)~10^{-10}$ is 
the pure electromagnetic contribution (see~\cite{hughes,cm} and references 
therein), \amuhadLO\ is the lowest-order contribution from hadronic 
vacuum polarization, $a_\mu^{\mathrm had,HO}=(-10.0\pm0.6)~10^{-10}$ 
is the corresponding higher-order part~\cite{krause2,adh}, 
and $a_\mu^{\mathrm weak}=(15.4\pm0.1\pm0.2)~10^{-10}$,
where the first error is the hadronic uncertainty and the second
is due to the Higgs mass range, accounts for corrections due to
exchange of the weakly interacting bosons up to two loops~\cite{amuweak}. 
For the LBL part we add the values for the pion-pole 
contribution~\cite{knecht_light,kino_light_cor,bij_light_cor} and the
other terms~\cite{kino_light_cor,bij_light_cor} to obtain
$a_\mu^{\mathrm had,LBL}=(8.6\pm3.5)~10^{-10}$.
\vs
By virtue of the analyticity of the 
vacuum polarization correlator, the contribution of the hadronic 
vacuum polarization to $a_\mu$ can be calculated \via\ the dispersion 
integral~\cite{rafael}
\beq\label{eq_int_amu}
    a_\mu^{\mathrm had,LO} \:=\: 
           \frac{\alpha^2(0)}{3\pi^2}
           \intl_{4m_\pi^2}^\infty\!\!ds\,\frac{K(s)}{s}R(s)~,
\eeq
where $K(s)$ is the QED kernel~\cite{rafael2}~,
\beq
      K(s) \:=\: x^2\left(1-\frac{x^2}{2}\right) \,+\,
                 (1+x)^2\left(1+\frac{1}{x^2}\right)
                      \left({\mathrm ln}(1+x)-x+\frac{x^2}{2}\right) \,+\,
                 \frac{(1+x)}{(1-x)}x^2\,{\mathrm ln}x~,
\eeq
with $x=(1-\beta_\mu)/(1+\beta_\mu)$ and $\beta_\mu=(1-4m_\mu^2/s)^{1/2}$.
In Eq.~(\ref{eq_int_amu}), $R(s)\equiv R^{(0)}(s)$ 
denotes the ratio of the 'bare' cross
section for \ee\ annihilation into hadrons to the pointlike muon-pair cross
section. The 'bare' cross section is defined as the measured cross section,
corrected for initial state radiation, electron-vertex loop contributions
and vacuum polarization effects in the photon propagator 
(see Section~\ref{sec_rad} for details). The reason for using the 'bare'
({\it i.e.} lowest order) cross section is that a full treatment of higher
orders is anyhow needed at the level of $a_\mu$, so that the use of 'dressed' 
cross sections would entail the risk of double-counting some of the 
higher-order contributions.
\vs
The function $K(s)$ decreases monotonically with increasing $s$. It gives
a strong weight to the low energy part of the integral~(\ref{eq_int_amu}).
About 91\pc\ of the total contribution to \amuhadLO\ 
is accumulated at center-of-mass 
energies $\sqrt{s}$ below 1.8~GeV and 73\pc\ of \amuhadLO\ is covered by 
the two-pion final state which is dominated by the $\rho(770)$ 
resonance. 

%
%

\section{The Input Data}
\label{sec_data}

%
%
\subsection{\it \ee\ Annihilation Data}
\label{sec_dat_ee}

The exclusive low energy \ee\ cross sections have been mainly measured by 
experiments running at \ee\ colliders in Novosibirsk and Orsay. Due to the 
high hadron multiplicity at energies above $\sim 2.5$~GeV, the exclusive 
measurement of the respective hadronic final states is not practicable.
Consequently, the experiments at the high energy colliders ADONE, SPEAR, 
DORIS, PETRA, PEP, VEPP-4, CESR and BEPC have measured 
the total inclusive cross section ratio 
$R$.
\vs
We give in the following a compilation of the data used in this analysis:
\begin{itemize}
\item The \ee$\rightarrow\pi^+\pi^-$ measurements are taken from 
  OLYA~\cite{barkov,E_54}, TOF~\cite{E_55}, CMD~\cite{barkov}, 
  DM1~\cite{E_58} and DM2~\cite{E_59}.
 
  The most precise data from CMD-2 are now available in their final 
  form~\cite{cmd2}. They differ from the preliminary ones, released 
  two years ago~\cite{cmd2_prel}, mostly in the treatment of radiative
  corrections. Compared to the preliminary ones, the new results are 
  corrected (see Section~\ref{sec_rad}) for leptonic 
  and hadronic vacuum polarization, and for photon radiation 
  by the pions (final state radiation -- FSR), 
  so that the measured final state corresponds to 
  $\pi^+\pi^-$ including pion-radiated photons. The various changes resulted 
  into a reduction of the cross section by about 1\% below the 
  $\rho$ peak and 5\% above. The dominant contribution stemmed from 
  vacuum polarization, while the (included) FSR correction 
  increased the cross section by about 0.8\% in the peak region. The overall
  systematic error of the final data is quoted to be 0.6\% and is
  dominated by the uncertainties in the radiative corrections (0.4\%).
 
  We do not use the data from NA7~\cite{E_56} as they are known 
  to suffer from a systematic bias in the energy scale~\cite{rolandi}. 
  All the experiments agree with each other within their quoted errors, 
  but the high precision claimed by CMD-2 makes this experiment unique 
  and consequently not cross-checked by the others at that level.

  The comparison between the cross section results from CMD-2 and from 
  previous experiments (corrected for vacuum polarization and FSR,
  according to the procedure discussed in Section~\ref{sec_rad})
  is shown in Fig.~\ref{fig_2pi_ee}. Note that the errors bars given 
  contain both statistical and systematic errors, added in quadrature 
  (this is the case for all figures in this paper).
  The agreement is
  good within the much larger uncertainties (2-10\%) quoted by the 
  older experiments. 

\begin{figure}[p]
\epsfxsize16cm
\centerline{\epsffile{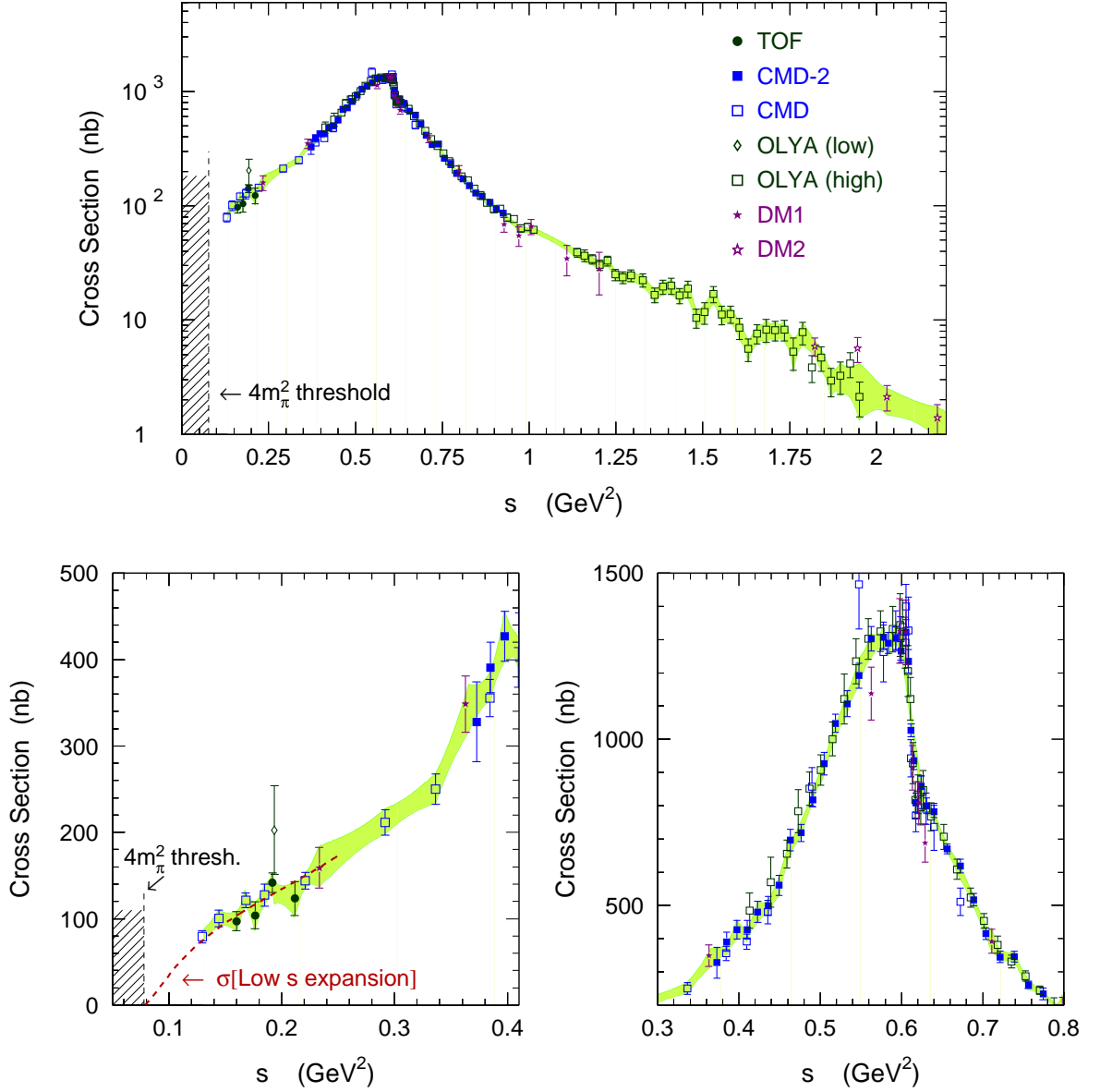}}
\caption[.]{\it The cross section for $e^+e^-\rightarrow\pi^+\pi^-~(\gamma)$
   measured by the different experiments. The errors bars contain
   both statistical and systematic errors, added in quadrature.
   The band is the combination
   of all the measurements used for the numerical integration following 
   the procedure discussed in Section~\ref{sec_integration}.}
\label{fig_2pi_ee}
\end{figure}

\item The situation of the data on the $\omega$ and $\phi$ resonances
  has significantly improved recently~\cite{snd_phi,cmd2_om,cmd2_phi}. The
  numerical procedure for integrating their cross sections is described
  in detail in Section~\ref{sec_omphi_int}.

\item The cross sections for \ee$\rightarrow\pi^0\gamma$ and $\eta\gamma$
  not originating from the decay of the $\omega$ and $\phi$ resonances 
  are taken from SND~\cite{sndpigam} and CMD-2~\cite{cmd2pigam} data
  in the continuum. They include contributions from 
  $\rho\rightarrow\pi^0\gamma$ and $\rho\rightarrow\eta\gamma$.

\item The reaction \ee$\rightarrow\pi^+\pi^-\pi^0$ 
  is dominated by the $\omega$ 
  and $\phi$ intermediate resonances discussed above. The continuum
  data are taken from ND~\cite{E_64}, DM1~\cite{E_68}, DM2~\cite{E_69},
  SND~\cite{snd3pi} and CMD~\cite{cmd_3pi}. 

\item The \ee$\rightarrow\pi^+\pi^-\pi^0\pi^0$ data are available from 
  M3N~\cite{E_66}, OLYA~\cite{ol}, ND~\cite{E_64}, 
  DM2~\cite{E_74,E_74p,E_75}, 
  CMD-2~\cite{cmd2_2pi2pi0} and SND~\cite{snd_2pi2pi0}. It is fair to say
  that large discrepancies are observed between the different results, which 
  are probably related to problems in the calculation of the detection
  efficiency (the cross sections can be seen in Fig.~\ref{fig_2pi2pi0_eetau} 
  shown in Section~\ref{sec_compsf}).
  The efficiencies are small in general ($\sim 10-30$\%) and 
  are affected by uncertainties in the decay dynamics that is assumed 
  in the Monte Carlo simulation. One could expect the more recent experiments 
  (CMD-2 and SND) to be more reliable in this context because of specific 
  studies performed in order to identify the major decay processes involved.
  Accordingly we do not include the ND data in the analysis.

\item The reaction \ee$\rightarrow\omega\pi^0$ is mainly reconstructed in 
  the $\pi^+\pi^-\pi^0\pi^0$ final state and is thus
  already accounted for. It was studied by the collaborations 
  ND~\cite{E_64}, DM2~\cite{E_74p}, CMD-2~\cite{cmd2_2pi2pi0} and 
  SND~\cite{snd_2pi2pi0,snd_piom}. We use these
  cross section measurements to compute the contribution
  corresponding to the $\omega \rightarrow \pi^0 \gamma$ 
  decay mode.

\item The \ee$\rightarrow\pi^+\pi^-\pi^+\pi^-$ final state was studied 
  by the experiments OLYA~\cite{E_70},
  ND~\cite{E_64}, CMD~\cite{E_72}, 
  DM1~\cite{E_73,PhysLett81B}, DM2~\cite{E_74,E_74p,E_75},
  CMD-2~\cite{cmd2_2pi2pi0} and SND~\cite{snd_2pi2pi0}. 
  The experiments agree reasonably well within
  their quoted uncertainties (see Fig.~\ref{fig_4pi_eetau} in 
  Section~\ref{sec_compsf}).

\item The \ee$\rightarrow\pi^+\pi^-\pi^+\pi^-\pi^0$ data are taken 
  from M3N~\cite{E_66} and CMD~\cite{E_72}. It contains a contribution from
  the $\eta\pi^+\pi^-$ channel with $\eta\rightarrow\pi^+\pi^-\pi^0$
  which has to be treated separately because the $\eta$ decay violates
  isospin. The other five-pion mode
  \ee$\rightarrow\pi^+\pi^-3\pi^0$ is not measured, but can be  
  accounted for using the isospin relation
  $\sigma_{\pi^+\pi^-3\pi^0}=\sigma_{\pi^+\pi^-\pi^+\pi^-\pi^0}/2$. The
  relation is used after subtracting the $\eta\pi^+\pi^-$ contribution
  in the $\pi^+\pi^-\pi^+\pi^-\pi^0$ rate. Then the $\eta\pi^+\pi^-$ 
  contribution with $\eta\rightarrow 3\pi^0$ is added to obtain the full
  $\pi^+\pi^-3\pi^0$ rate.

\item For the reaction \ee$\rightarrow\omega\pi^+\pi^-$, measured by the 
  groups DM1~\cite{dm1_ompp}, DM2~\cite{E_69} and CMD-2~\cite{cmd2_ompp}, 
  a contribution is calculated
  for $\omega$ decaying into $\pi^0\gamma$. The dominant three-pion decay  
  already appears in the five-pion final state.

\item Similarly, the contribution for \ee$\rightarrow\omega2\pi^0$, with
  $\omega\rightarrow\pi^0\gamma$, is taken by isospin symmetry to be half
  of \ee$\rightarrow\omega\pi^+\pi^-$. 

\item The process \ee$\rightarrow\eta\pi^+\pi^-$ was studied by 
  ND~\cite{E_64}, DM2~\cite{E_69} and CMD-2~\cite{cmd2_ompp}.
  We subtract from its cross section the 
  contributions which are already counted in the $\pi^+\pi^-\pi^+\pi^-\pi^0$ 
  and $\pi^+\pi^-3\pi^0$ final states.

\item The cross sections of the six-pion final states $3\pi^+3\pi^-$ 
  and $2\pi^+2\pi^-2\pi^0$ were measured by DM1~\cite{E_79},
  CMD~\cite{E_72} and DM2~\cite{schioppa}. For the missing channel
  $\pi^+ \pi^- 4\pi^0$ one can rely on isospin relations in order to estimate
  its contribution. If only \ee\ data are used, the isospin bound~\cite{adh}
  is weak, leading to a possibly large contribution with an equally large
  uncertainty. However, some information can be found in the isospin-rotated
  processes\footnote{Throughout this paper, charge conjugate states are implied
  for $\tau$ decays.} $\tau^- \rightarrow \nu_\tau \,3\pi^- 2\pi^+ \pi^0$ 
  and $\tau^- \rightarrow \nu_\tau \,2\pi^- \pi^+ 3\pi^0$, where the hadronic 
  system has been shown~\cite{cleo_6pi} to be dominated by 
  $\omega 2\pi^- \pi^+$ and $\omega \pi^- 2\pi^0$, once the axial-vector
  $\eta 2\pi^- \pi^+$ and $\eta \pi^- 2\pi^0$ 
  contributions~\cite{cleoeta3pi} are discarded.
  An isospin analysis then reveals the dominance of the 
  $\omega \rho^\pm \pi^\mp$
  final state. As a consequence the $\pi^+ \pi^- 4\pi^0$ channel in \ee\
  annihilation only receives a very small contribution, determined by
  the $3\pi^+3\pi^-$ cross section. We include a component for    
  $\omega \rar \pi^0\gamma$.

\item The \ee$\rightarrow$\,$K^+K^-$ and \ee$\rightarrow$\,\Ks\Kl\ 
  cross sections above the $\phi$ resonance are taken from 
  OLYA~\cite{E_82}, DM1~\cite{LAL80_xx}, 
  DM2~\cite{E_87}, CMD~\cite{cmd_kk} and CMD-2~\cite{cmd2_kk}.

\item The reactions \ee$\rightarrow$\,\Ks\,$K^\pm\pi^\mp$ and  
  \ee$\rightarrow$\,$K^+K^-\pi^0$ were studied by DM1~\cite{E_88a,E_88b} 
  and DM2~\cite{E_74}. Using isospin symmetry the cross 
  section of the final state \Ks\Kl\piz\ is obtained from the relation 
  $\sigma_{K_{\mathrm S}^0K_{\mathrm L}^0\pi^0}
  =\sigma_{K^+K^-\pi^0}$.

\item The inclusive reaction \ee$\rightarrow$\,\Ks+$X$ was analyzed by 
  DM1~\cite{DM1thesis}. After subtracting from its cross section the separately
  measured contributions of the final states \Ks\Kl, \Ks\,$K^\pm\pi^\mp$ and 
  \Ks\Kl\piz, it still includes the modes \Ks\Ks$\pi^+\pi^-$, 
  \Ks\Kl$\pi^+\pi^-$ and 
  \Ks$K^\pm\pi^\mp\pi^0$.
  With the assumption that the cross sections for the processes 
  \ee$\,\rightarrow K^0\overline{K}^0(\pi\pi)^0$ and
  \ee$\,\rightarrow K^+K^-(\pi\pi)^0$ are equal, one can
  summarize the total $K\overline{K}\pi\pi$ contribution as twice 
  the above corrected \Ks+$X$ cross section. Implied by the 
  assumption made, it is reasonable to quote as the systematic uncertainty
  one-half of the cross section for the channel 
  $K^+K^-\pi^+\pi^-$ measured by DM1~\cite{E_88a} 
  and DM2~\cite{dm2_kkpp}.

\item Baryon-pair production is included using the cross sections
  for $p \overline{p}$ from DM1~\cite{dm1_pp} and DM2~\cite{dm2_pp}, and
  for $n \overline{n}$ from FENICE~\cite{fenice}.

\item At energies larger than 2~GeV the total cross section ratio 
  $R$ is measured inclusively. Data are provided by the experiments 
  $\gamma\gamma2$~\cite{E_78}, MARK~I~\cite{E_96}, DELCO~\cite{delco}, 
  DASP~\cite{dasp}, PLUTO~\cite{pluto}, LENA~\cite{lena}, 
  Crystal Ball~\cite{CB,CBcharm},
  MD-1~\cite{MD1}, CELLO~\cite{cello}, JADE~\cite{jade}, MARK-J~\cite{markj},
  TASSO~\cite{tasso}, CLEO~\cite{cleo}, CUSB~\cite{cusb}, MAC~\cite{mac},
  and BES~\cite{bes}. 
  Due to their weak experimental precision, the data of $\gamma\gamma2$
  are not used in this analysis.
  The measurements of the MARK~I Collaboration are significantly 
  higher than those from more recent and more precise experiments.
  In addition, the QCD prediction of $R$, which should be reliable in this 
  energy regime, favours lower values, in agreement with the other
  experiments. Consequently the MARK~I results on $R$ have been discarded. 

  Although small, the enhancement of the cross section due to 
  $\gamma-Z$ interference is corrected for energies above the $J/\psi$ 
  mass. We use a factorial ansatz according to Ref.~\cite{burkjeger,eidelman},
  yielding a negligible contribution to \amuhadLO.

  The $R$ data in the charm region are displayed in Fig.~\ref{fig_r_cc}. 
  Good agreement is found among the experiments.

\begin{figure}[t]
\epsfxsize12cm
\centerline{\epsffile{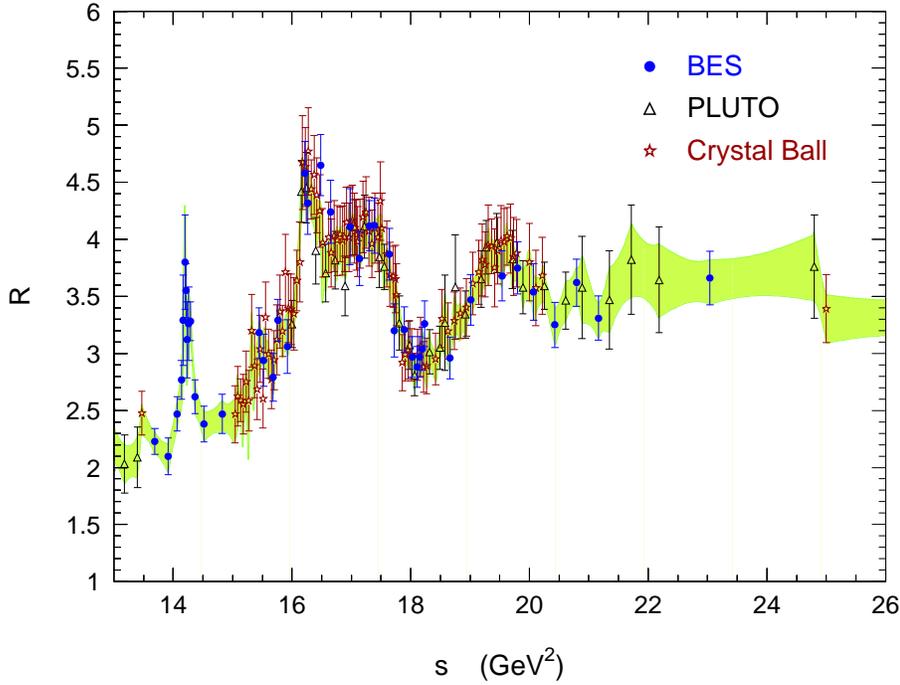}}
\caption[.]{\it $R$ data in the charm region. The band is the combination
   of all the measurements used for the numerical integration following 
   the procedure discussed in Section~\ref{sec_integration}.}
\label{fig_r_cc}
\end{figure}

\item The narrow $c\bar{c}$ and $b\bar{b}$ resonances are treated in
Section~\ref{sec_psi}.

\end{itemize}

%
%
\subsection{\it Data from Hadronic $\tau$ Decays}

Data from $\tau$ decays into two- and four-pion final states
\tauto\nut\pipiz, \tauto\nut\pitpiz\ and \tauto\nut\tpipiz,
are available from ALEPH~\cite{aleph_vsf}, 
CLEO~\cite{cleo_2pi,cleo_4pi} and OPAL~\cite{opal}.
Very recently, preliminary results on the full LEP1 
statistics have been presented by ALEPH~\cite{aleph_new}. They agree 
with the published results, but correspond to a complete reanalysis
with refined systematic studies allowed by the 2.5 times larger data set.  
The branching fraction $B_{\pi\pi^0}$ for the 
$\tau\rightarrow\nu_\tau\,\pi^-\pi^0~(\gamma)$ decay mode is of 
particular interest since it provides the normalization of the
corresponding spectral function. The new value~\cite{aleph_new},
$B_{\pi\pi^0}=(25.47 \pm 0.13)~\%$, turns out to be larger
than the previously published one~\cite{aleph_h} based on the 1991-93
LEP1 statistics, $(25.30 \pm 0.20)~\%$.
\vs
Assuming (for the moment) isospin invariance to hold, 
the corresponding \ee\ isovector
cross sections are calculated \via\ the CVC relations
\begin{eqnarray}
\label{eq_cvc_2pi}
 \sigma_{e^+e^-\rightarrow\,\pi^+\pi^-}^{I=1}
        & \:=\: &
 \frac{4\pi\alpha^2}{s}\,v_{\pi^-\pi^0}~, \\[0.3cm]
\label{eq_cvc_4pi}
 \sigma_{e^+e^-\rightarrow\,\pi^+\pi^-\pi^+\pi^-}^{I=1} 
        & \:=\: &
             2\cdot\frac{4\pi\alpha^2}{s}\,
             v_{\pi^-\,3\pi^0}~, \\[0.3cm]
\label{eq_cvc_2pi2pi0}
 \sigma_{e^+e^-\rightarrow\,\pi^+\pi^-\pi^0\pi^0}^{I=1} 
        & \:=\: &
             \frac{4\pi\alpha^2}{s}\,
             \left[v_{2\pi^-\pi^+\pi^0} 
                  \:-\:
                     v_{\pi^-\,3\pi^0}
             \right]~.
\end{eqnarray}

\begin{figure}[p]
\epsfxsize16cm
\centerline{\epsffile{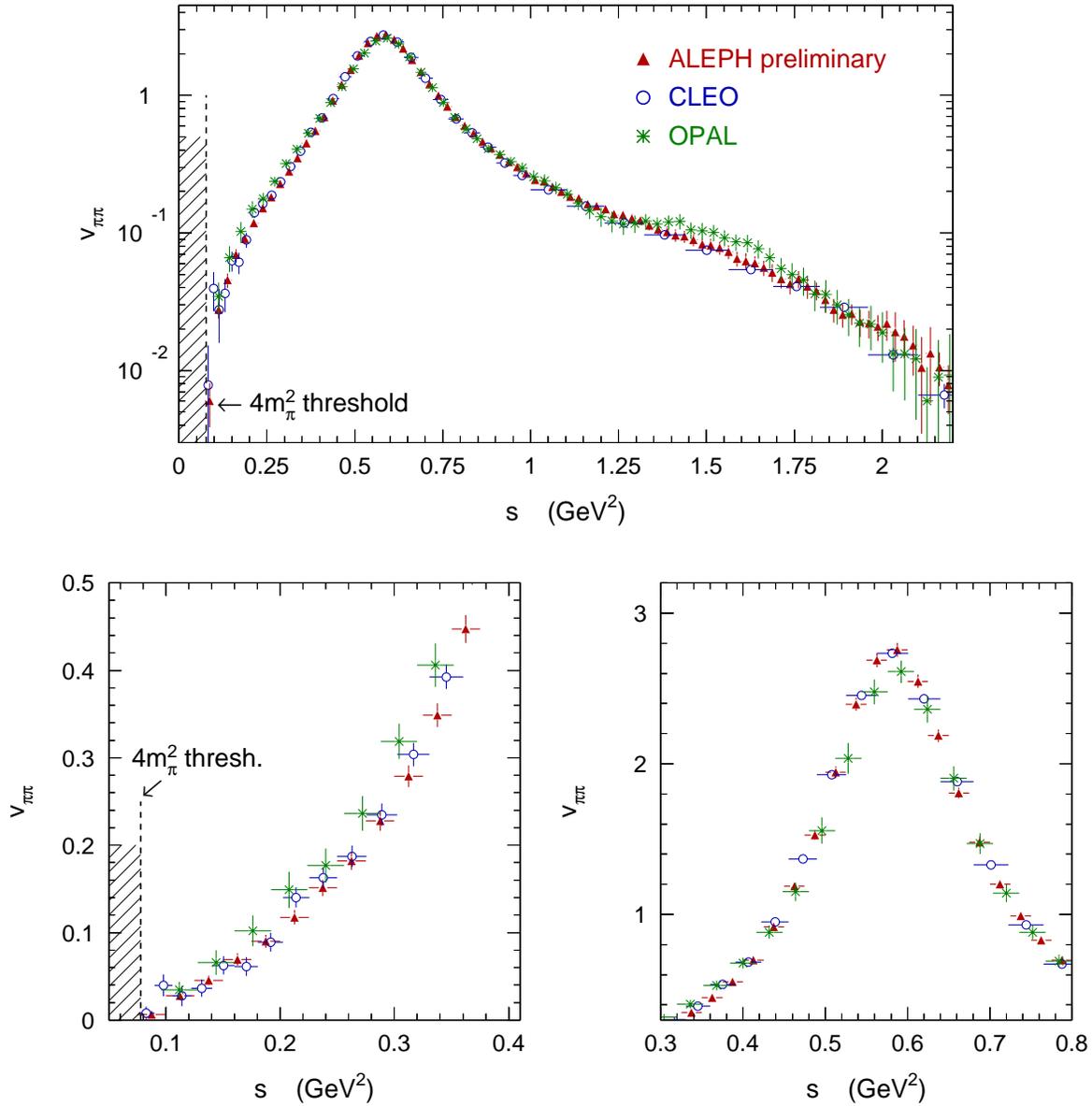}}
\caption[.]{\it Comparison between the shapes of the ALEPH, CLEO and OPAL
      $\pi\pi$ \sfs\  normalized to the world-average
      branching ratio $B_{\pi\pi^0}$. The normalization errors,
      correlated between the shapes, are not contained in the 
      error bars.  }
\label{fig_2pi_tau}
\end{figure}

The $\tau$ \sf\ $v_V(s)$ for a given vector hadronic state $V$ is 
defined by~\cite{tsai}
\beq
\label{eq_sf}
   v_V(s) 
   \equiv
           \frac{m_\tau^2}{6\,|V_{ud}|^2\,S_{\mathrm{EW}}}\,
              \frac{B(\tau^-\rightarrow \nu_\tau\,V^-)}
                   {B(\tau^-\rightarrow \nu_\tau\,e^-\,\bar{\nu}_e)}         
              \frac{d N_{V}}{N_{V}\,ds}\,
              \left[ \left(1-\frac{s}{m_\tau^2}\right)^{\!\!2}\,
                     \left(1+\frac{2s}{m_\tau^2}\right) \right]^{-1},
\eeq
where $|V_{ud}|=0.9748\pm0.0010$ is obtained from averaging\footnote
{
	Since the two determinations, 
	$|V_{ud}|_{\rm nucleons}=0.9734\pm0.0008$ and
	$|V_{ud}|_{\rm kaons}=0.9756\pm0.0006$ are not consistent, 
	the final error has been enlarged correspondingly.
} the two
independent determinations~\cite{pdg2002} from nuclear $\beta$ decays 
and kaon decays (assuming unitarity of the CKM matrix) and $S_{\mathrm{EW}}$ 
accounts for electroweak radiative corrections as discussed 
in Section~\ref{sec_isobreak1}.
The \sfs\ are obtained from the corresponding invariant mass distributions,
after subtracting out the non-$\tau$ background and the feedthrough from 
other $\tau$ decay channels, and after a final unfolding from 
detector effects such as energy and angular resolutions, acceptance,
calibration and photon identification.
\vs
It is important to note that $\tau$ decay experiments measure decay rates 
that include the possibility of photon radiation in the decay final state.
Depending on the experiment, the analysis may (ALEPH) or may not  
(CLEO) keep events with radiative photons in the final state, but 
all experiments rely on the
TAUOLA $\tau$ decay library~\cite{was} to compute their efficiencies. In
TAUOLA charged particles are given a probability to
produce bremsstrahlung using the PHOTOS procedure~\cite{photos}
which is based on the leading logarithm approximation valid at low 
photon energy.
Thus the measured \sfs\ correspond to given final states inclusive with 
respect to radiative photons in the $\tau$ decay.
\vs
It should be pointed out that the
experimental conditions at $Z$ (ALEPH, OPAL) and $\Upsilon (4S)$ (CLEO)
energies are very different. On the one hand, at LEP, 
the $\tau^+\tau^-$ events can be
selected with high efficiency ($>90\%$) and small non-$\tau$ 
background ($<1\%$), thus ensuring little bias in the efficiency
determination. The situation is not as favorable at lower energy: because 
the dominant hadronic cross section has a smaller particle multiplicity,
it is more likely to pollute the $\tau$ sample and strong cuts must be 
applied, hence resulting in smaller efficiencies. On the other hand,
CLEO has an advantage for the reconstruction of the decay final 
state since particles are more separated in space. The LEP detectors
have to cope with collimated $\tau$ decay products and the granularity
of the detectors, particularly the calorimeters, plays a crucial role.
One can therefore consider ALEPH/OPAL and CLEO data to be approximately 
uncorrelated as far as experimental procedures are concerned.
The fact that their respective spectral functions for 
the $\pi^-\pi^0$ and $2\pi^-\pi^+\pi^0$ modes agree, as demonstrated
in Fig.~\ref{fig_2pi_tau} for $\pi^-\pi^0$, is therefore a valuable 
experimental consistency test.

%
%
\section{Radiative Corrections for \boldmath\ee\ Data}
\label{sec_rad}

Radiative corrections applied to the measured \ee\ cross sections are an
important step in the experimental analyses. They involve the consideration
of several physical processes and lead to large corrections. We stress
again that the evaluation of the integral in Eq.~(\ref{eq_int_amu}) requires
the use of the 'bare' hadronic cross section, so that the input data must
be analyzed with care in this respect.
\vs
Several steps are to be considered in the radiative correction procedure:
\bei
\item Corrections are applied to the luminosity determination, based
on large-angle Bhabha scattering and muon-pair production in the low-energy
experiments, and small-angle Bhabha scattering at high energies. These
processes are usually corrected for external radiation, vertex corrections
and vacuum polarization from lepton loops.
\item The hadronic cross sections given by the experiments are always 
corrected for initial state radiation and the effect of loops 
at the electron vertex.
\item The vacuum polarization correction in the photon propagator is
a more delicate point. The cross sections need to be fully corrected 
for our use, {\it i.e.}
\beq
\label{eq_dress}
  \sigma_{\rm bare} \,=\, \sigma_{\rm dressed}
          \left(\frac{\alpha(0)}{\alpha(s)}\right)^{\!2} ~,
\eeq
where $\sigma_{\rm dressed}$ is the measured cross section 
already corrected for initial state radiation, 
and $\alpha(s)$ is obtained from resummation of the lowest-order
evaluation
\beq
\label{eq_alpha}
  \alpha(s) \,=\, \frac{\alpha(0)}
                 {1 -\Delta\alpha_{\rm lep}(s) -\Delta\alpha_{\rm had}(s)}.
\eeq
Whereas $\Delta\alpha_{\rm lep}(s)$ can be analytically calculated (here
given to leading order)
\beq
\label{eq_dalpha_lep}
  \Delta\alpha_{\rm lep}(s)\,=\,\frac{\alpha(0)}{3\pi} \sum_l 
                      \left(\log \frac{s}{m_l^2} -\frac{5}{3}\right)~,
\eeq
$\Delta\alpha_{\rm had}(s)$ is related by analyticity and unitarity to a
dispersion integral, akin to~(\ref{eq_int_amu}),
\beq
\label{eq_dalpha_had}
  \Delta\alpha_{\rm had}(s) \:=\:
        -\frac{\alpha(0) s}{3\pi}\,
         {\rm Re}\!\!\intl_{4m_\pi^2}^{\infty}\!\!ds^\prime\,
            \frac{R(s^\prime)}{s^\prime(s^\prime-s-i\epsilon)}~,
\eeq
which must also be evaluated using input data. Since the hadronic correction
involves the knowledge of $R(s)$ at all energies, including
those where the measurements are made, the procedure has to be iterative,
and requires experimental as well as theoretical information  
over a large energy range.

This may explain why the vacuum polarization correction is in general 
not applied by the experiments to their published cross sections. 
Here the main difficulty is even to find out
whether the correction (and which one? leptonic at least? hadronic?) has
actually been used, as unfortunately this is almost never clearly stated 
in the publications. The new data from CMD-2~\cite{cmd2} are explicitly
corrected for both leptonic and hadronic vacuum polarization effects,
whereas the preliminary data from the same experiment~\cite{cmd2_prel} 
were not. 

In fact, what really matters is the correction to the ratio of the hadronic
cross section to the cross section for the process used for the luminosity
determination. In the simplest case (for example, DM2 for the $\pi^+\pi^-$
channel) of the normalization to the $e^+e^-\rar\mu^+\mu^-$ process, the 
vacuum polarization effects cancel. 
However, generally the normalization is done with respect
to large angle Bhabha scattering 
events or to both Bhabha and $\mu^+\mu^-$. In the latter
case, Bhabha events dominate due to the $t$-channel contribution. In the
$\pi^+\pi^-$ mode, all experiments before the latest CMD-2 results 
corrected their measured processes ($\pi^+\pi^-$, $\mu^+\mu^-$ and $e^+e^-$)
for radiative effects using $O(\alpha^3)$ calculations which took only
leptonic vacuum polarization into account~\cite{bonneau_rad,eidelman_rad}.
For the other channels, it is harder to find out as information about the 
luminosity determination and the detailed procedure for radiative 
corrections is in general not given in the publications.

For all \ee\ experimental results, but the newest $\pi^+\pi^-$ from CMD-2 
and DM2, we apply a correction $C_{\rm HVP}$ for the missing hadronic 
vacuum polarization given by~\cite{swartz}
\beq
 C_{\rm HVP}\,=\,\frac{1-2\Delta\alpha_{\rm had}(s)}
            {1-2\Delta\alpha_{\rm had}(\overline{t})}~,
\eeq
where the correction in the denominator applies to the Bhabha cross section
evaluated at a mean value of the squared momentum transfer $t$, which depends
on the angular acceptance in each experiment. A $50\%$ uncertainty is
assigned to $C_{\rm HVP}$. For the $\omega$ and $\phi$ resonance 
cross sections, we were informed that
the recent CMD-2 and SND results were not corrected for leptonic vacuum
polarization, so in their case we applied a full correction taking into
account both leptonic and hadronic components.

\item In Eq.~(\ref{eq_int_amu}) one must incorporate in $R(s)$ the
contributions of all hadronic states produced at the energy $\sqrt{s}$.
In particular, radiative effects in the hadronic final state
must be considered, \ie, final states such as $V+\gamma$ have 
to be included. 

Investigating the existing data in this respect is also a difficult task. 
In the $\pi^+\pi^-$ data from CMD-2~\cite{cmd2} most additional photons 
are experimentally rejected to reduce backgrounds from other channels and 
the fraction kept is subtracted using the Monte Carlo simulation 
which includes a model for FSR. Then the full FSR contribution is
added back as a correction, using an analytical expression computed in
scalar QED (point-like pions)~\cite{jeger_rad}. As this effect was not 
included in earlier analyses, we applied the same correction 
to older $\pi^+\pi^-$ data.

In principle one must worry about FSR effects in other channels as well. 
For the inclusive $R$ measurements it is included by definition. When $R$ 
is evaluated from QCD at high energy, the prediction must be corrected for
FSR from the quarks, but this is a negligible effect for \amuhadLO. The
situation for the exclusive channels is less clear because it
depends on the experimental cuts and whether or not FSR is included in
the simulation. Taking as an educated guess the effect in the
$\pi^+\pi^-$ channel, we correspondingly correct the contributions
to \amuhadLO\ from all remaining exclusive channels by 
the factor $C_{\rm FSR}=(1.004\pm0.004)^{n_c}$ where $n_c$ is the 
charged particle multiplicity in the final state. 
\eei

In summary, we correct each \ee\ experimental result, but those from
CMD-2 $\pi\pi$, by the factor $C_{\rm rad}=C_{\rm HVP}C_{\rm FSR}$. 
As an illustration of the orders of magnitude involved, the
different corrections in the $\pi^+\pi^-$ contribution amount 
to $-2.3\%$ for the leptonic vacuum polarization, $+0.9\%$ for the 
hadronic vacuum polarization, and $+0.9\%$ for the FSR correction.
The correction to the $\pi\pi$/$ee$ ratio from
the missing hadronic vacuum polarization is small, typically $0.56\%$.
Both the vacuum polarization and FSR corrections apply only to experiments
other than CMD-2, therefore the overall correction to the $\pi\pi$
channel is considerably reduced.
\vs
The uncertainties on the missing vacuum polarization ($50\%$) and 
the FSR corrections ($100\%$) are conservatively considered to be 
fully correlated between all channels to which the correction applies.
The total error from these missing radiative corrections, taken as 
the quadratic sum of the two contributions, is given separately for 
the final results.

%
%
\section{Isospin Breaking in \boldmath\ee\ and \boldmath$\tau$ 
	 Spectral Functions}
\label{sec_cvc}

%
%
\subsection{\em Sources of Isospin Symmetry Breaking}
\label{sec_isobreak1}

The relationships~(\ref{eq_cvc_2pi}), (\ref{eq_cvc_4pi}) and 
(\ref{eq_cvc_2pi2pi0}) between \ee\ and $\tau$ spectral functions only hold
in the limit of exact isospin invariance. This is the Conserved Vector 
Current (CVC) property of weak decays. It follows from the factorization
of strong interaction physics as produced through the $\gamma$ and $W$
propagators out of the QCD vacuum.
However, we know that we must expect symmetry 
breaking at some level from electromagnetic effects and even in QCD  
because of the up and down quark mass splitting. Since the normalization
of the $\tau$ spectral functions is experimentally known at the 0.5\%
level, it is clear that isospin-breaking effects must be carefully examined
if one wants this precision to be maintained in the vacuum polarization 
integrals. Various identified sources of isospin breaking are 
considered in this section and discussed in turn.
\vs
Because of the dominance of the $\pi\pi$ contribution in the energy range 
of interest for $\tau$ data, we discuss mainly this channel,
following our earlier analysis~\cite{adh}. The corrections on \amuhadLO\
from isospin breaking are given in Table~\ref{tab_isobreak}. 
A more complete evaluation is given in the next section. 
Finally, the 4-pion modes will be briefly discussed.
 
\bei
\item Electroweak radiative corrections must be taken into account. 
Their dominant contribution comes from the short distance correction
to the effective four-fermion coupling 
$\tau^-\rightarrow\nu_\tau(d\bar{u})^-$
enhancing the $\tau$ amplitude by the factor 
$(1+3\alpha(m_\tau)/4\pi)(1+2\overline{Q})\,{\mathrm ln}
\left(M_{Z}/m_\tau\right)$, where $\overline{Q}$ is the
average charge of the final state partons~\cite{marciano-sirlin}. While 
this correction vanishes for leptonic decays, it contributes for quarks.
All higher-order logarithms can be resummed using the renormalization
group~\cite{marciano-sirlin,bnp}, and the short distance 
correction can be absorbed into an overall multiplicative electroweak 
correction $S^{\rm had}_{\mathrm{EW}}$,
\beq
  S^{\rm had}_{\mathrm{EW}}\,=\,
     \left(\frac{\alpha(m_b)}{\alpha(m_\tau)}\right)^{\!9/19}
     \left(\frac{\alpha(M_W)}{\alpha(m_b)}\right)^{\!9/20}
     \left(\frac{\alpha(M_Z)}{\alpha(M_W)}\right)^{\!36/17},
\eeq
which is equal to 1.0194 when using the current fermion and boson masses
and for consistency~\cite{sew_private} the quark-level $\overline{\rm MS}$
expressions for $\alpha(s)$ as given in Ref.~\cite{marciano-sirlin2}. 
The difference between the resummed value and the lowest-order 
estimate (1.0188) can be taken as a conservative estimate of the uncertainty. 
QCD corrections to $S^{\rm had}_{\mathrm{EW}}$ have been 
calculated~\cite{marciano-sirlin, sirlin2} and found to be small, 
reducing its value to 1.0189.

Subleading non-logarithmic short distance 
corrections have been calculated to order $O(\alpha)$ at the quark 
level~\cite{braaten}, 
$S^{\rm sub,had}_{\mathrm{EW}}=1+\alpha(m_\tau)(85/12-\pi^2)/(2\pi)\simeq0.9967$, 
and for the leptonic width~\cite{marciano-sirlin}, 
$S^{\rm sub,lep}_{\mathrm{EW}}=1+\alpha(m_\tau)(25/4-\pi^2)/(2\pi)\simeq0.9957$.
Summing up all the short distance corrections, one obtains the value for
$S_{\mathrm{EW}}$ that must be used for the inclusive hadronic width
\beq
\label{sew}
    S_{\mathrm{EW}}^{\rm inclusive}=\frac{S^{\rm had}_{\mathrm{EW}}\,
           S^{\rm sub,had}_{\mathrm{EW}}}{S^{\rm sub,lep}_{\mathrm{EW}}}
                    =1.0199\pm0.0006~.
\eeq
Other uncertainties on the $b$ quark mass, the running of $\alpha(s)$, 
and QCD corrections are at the $10^{-4}$ level. 

Long distance corrections are expected to be final-state dependent
in general. They have been computed for the \tauto$\nu_\tau\pi^-$ 
decay leading to a total radiative correction of 2.03\pc~\cite{decker}, 
which is dominated by the leading logarithm from the short distance 
contribution. Although very encouraging, this result may not apply to
all hadronic $\tau$ decays, in particular for the important 
$\nu_\tau\pi^-\pi^0$ mode. Therefore
an uncertainty of 0.0040 was previously assigned to $S_{\mathrm{EW}}$
(see the following Section~\ref{sec_isobreak2} for more) to cover
the final-state dependence of the correction with respect to the
calculation at the quark level. 
\item A contribution~\cite{czyz,adh} for isospin breaking occurs 
because of the mass difference between charged and neutral pions, which
is essentially of electromagnetic origin. The \sf\ has a kinematic factor
$\beta^3$ which is different in \ee\ ($\pi^+\pi^-$) and $\tau$ decay
($\pi^-\pi^0$). We write
\beqn
\label{eq_v0}
 v_0(s) & = &\frac{\beta_0^3(s)}{12}|F_\pi^0(s)|^2~, \\
\label{eq_v-}
 v_-(s) & = &\frac{\beta_-^3(s)}{12}|F_\pi^-(s)|^2~,
\eeqn
with obvious notations, $F_\pi^{0,-}(s)$ being the electromagnetic and 
weak pion form factors, respectively, and $\beta_{0,-}$ defined by
\beq
 \beta_{0,-}=\beta(s,m_{\pi^-},m_{\pi^{0,-}})~,
\eeq
where
\beq
 \beta(s,m_1,m_2)=\left[\left(1-\frac{(m_1+m_2)^2}{s}\right)
                     \left(1-\frac{(m_1-m_2)^2}{s}\right)\right]^{1/2}~.
\eeq
Hence, a correction equal to $\beta^3_0(s)/\beta^3_-(s)$
is applied to the $\tau$ \sf.
\item Other corrections occur in the form factor itself. It turns out 
that it is affected by the pion mass difference because the same
$\beta^3$ factor enters in the $\rho\rightarrow\pi\pi$ width. This effect
partially compensates the $\beta^3$ corrections~(\ref{eq_v0}), (\ref{eq_v-})
of the cross section, as seen in Table~\ref{tab_isobreak}.
\item Similarly a possible 
mass difference between the charged and neutral $\rho$
meson affects the value of the corresponding width and
shifts the resonance lineshape. Theoretical estimates~\cite{bijnens} 
and experimental determinations~\cite{aleph_vsf,snd_rho} show that the mass
difference is compatible with zero within about 1~MeV.
\item $\rho-\omega$ interference occurs in the $\pi^+\pi^-$ mode
and thus represents an obvious source of isospin symmetry breaking. Its
contribution can be readily introduced into the $\tau$ \sf\ using the
parameters determined in the CMD-2 fit~\cite{cmd2}. The integral over
the interference  almost vanishes by itself since it changes sign 
at the $\omega$ mass, however the $s$-dependent integration kernel 
produces a net effect (Table~\ref{tab_isobreak}).
\item Electromagnetic $\rho$ decays explicitly break SU(2) symmetry. 
This is the case for the decays $\rho\rightarrow\pi\pi^0\gamma$
through an $\omega$ intermediate state because of identical $\pi^0$'s,
$\rho\rightarrow\pi\gamma$, $\rho^0\rightarrow\eta\gamma$ and 
$\rho^0\rightarrow l^+l^-$. The decay $\rho\rightarrow\pi\pi\gamma$
deserves particular attention: calculations have been done with an
effective model~\cite{singer} for both charged and neutral $\rho$'s.
The different contributions are listed in Table~\ref{tab_isobreak}.
\item A breakdown of CVC is due to quark mass effects: 
$m_u$ different from $m_d$ generates
$\partial_{\mu}J^\mu\sim(m_u - m_d)$ for a charge-changing hadronic 
current $J^\mu$ between $u$ and $d$ quarks.
Expected deviations from CVC 
due to so-called second class currents such as, {\it e.g.}, the decay 
$\tau^-\rightarrow\nu_\tau\pi^-\eta$ where the corresponding \ee\ final state
\piz$\eta$ (C=+1) is forbidden, lead to an estimated branching 
fraction of the order of $(m_u-m_d)^2/m_\tau^2\simeq10^{-5}$~\cite{etapi}, 
while the experimental upper limit amounts to 
$B(\tau\rightarrow\nu_\tau\pi^-\eta)$\,$\,<1.4~10^{-4}$~\cite{pdg2002}.
\eei

\begin{table}[t]
\begin{center}
\setlength{\tabcolsep}{0.35pc}
\begin{tabular}{lcccc} \hline
&&&\\[-0.3cm]
Sources of Isospin 	& \mc{4}{c}{$\Delta$\amuhadLO~($10^{-10}$)}      \\
Symmetry Breaking 
	& $\pi^+\pi^-$ (I)
		& $\pi^+\pi^-$ (II) 
			& $\pi^+\pi^- 2\pi^0$ & $2\pi^+ 2\pi^-$ \\[0.15cm]
\hline
&&&&\\[-0.3cm]
Short distance rad. corr.
	& 	& $-12.1\pm0.3 $ &         &        \\
Long distance rad. corr.
	& \rs{$-10.3 \pm 2.1$} & $-1.0 $ & 
    \rs{$-0.36 \pm 0.07$}  & \rs{$-0.18 \pm 0.04$}   \\
$m_{\pi^-}\ne m_{\pi^0}$ ($\beta$ in cross section)
	& $-7.0$ & $-7.0$ &  $+0.6 \pm 0.6$ & $-0.4 \pm 0.4$            \\ 
$m_{\pi^-}\ne m_{\pi^0}$ ($\beta$ in $\rho$ width)    
	& $+4.2$ & $+4.2$ & --     & --             \\ 
$m_{\rho^-}\ne m_{\rho^0}$               
	& $0 \pm 0.2$ & $0 \pm 2.0$ & --   & --           \\
$\rho-\omega$ interference              
	& $+3.5 \pm 0.6$ & $+3.5 \pm 0.6$ & --  & --         \\
Electromagnetic decay modes                   
	& $-1.4 \pm 1.2$ & $-1.4 \pm 1.2$ & -- & --  \\[0.15cm]
\hline 
&&&\\[-0.3cm]
Sum                                     
	& $-11.0 \pm 2.5$ & $-13.8 \pm 2.4$ & 
       $+0.2 \pm 0.6$ & $-0.6 \pm 0.4$ \\[0.15cm]
\hline
\end{tabular}
\end{center}
\caption{\label{tab_isobreak}\it
         Expected sources of isospin symmetry breaking between 
         \ee\ and $\tau$ \sfs\ in the $2\pi$ and $4\pi$ channels, 
         and the corresponding corrections
         to \amuhadLO\ as obtained from $\tau$ data. The corrections (I)
         follow essentially the procedure used in Refs.~\cite{adh,dh97,dh98},
         while in (II) the more complete approach of Ref.~\cite{ecker2} is
         chosen. The values
         given for (II) differ slightly from those quoted
         in Ref.~\cite{ecker2}, because of the model 
	 used in the latter to parametrize the pion form factor, in addition
         to the re-evaluation of the short distance electroweak correction.
	 The errors given are theoretical only. Uncertainties introduced
	 by the experimental error on the $\tau$ spectral function itself
	 are not accounted for here.}
\end{table} 

%
%
\subsection{\it A More Elaborate Treatment of Isospin Breaking 
in the $2\pi$ Channel}
\label{sec_isobreak2}

\begin{figure}[t]
\epsfxsize12cm
\centerline{\epsffile{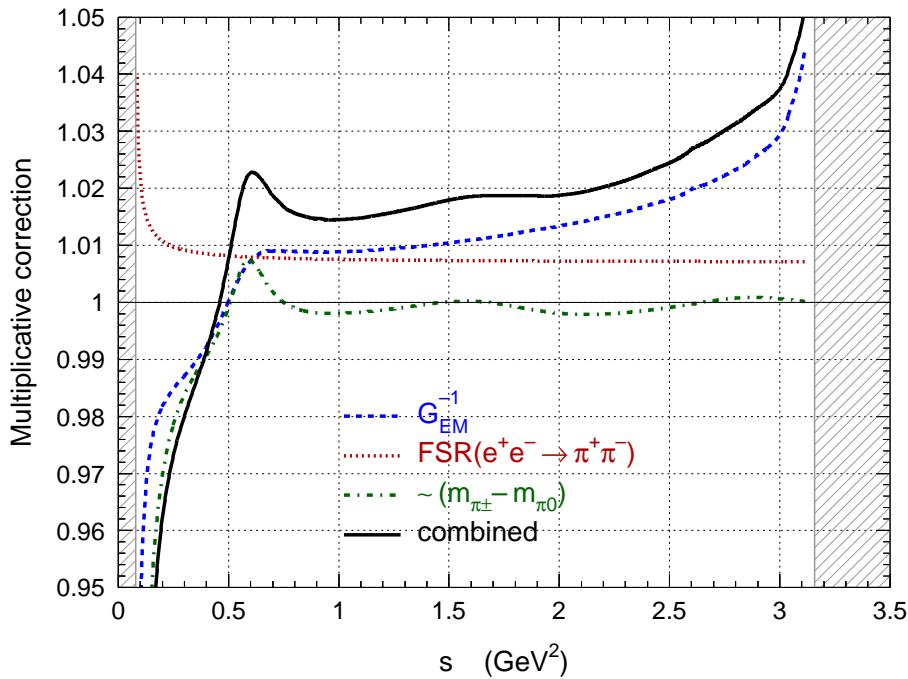}}
\caption[.]{\it Mass-squared-dependent corrections applied to
          the $\pi^-\pi^0$ spectral function from $\tau$ data,
          following the analysis of Ref.~\cite{ecker2}.}
\label{fig_isobreak}
\end{figure}

The above analysis of isospin breaking leaves out the possibility of
sizeable contributions from virtual loops. This problem was studied
recently~\cite{ecker1} within a model based on Chiral Perturbation Theory.
In this way the correct low-energy hadronic structure is implemented and
a consistent framework can be set up to calculate electroweak and strong
processes, such as the radiative corrections in the 
$\tau\rightarrow\nu_\tau\pi^-\pi^0$ decay. One might worry that the $\rho$
mass is too large for such a low-energy approach. However a
reasonable matching with the resonance region~\cite{pich_fpi} 
and even beyond is claimed to be achieved, providing a very useful 
tool to study radiative decays.
\vs
A new analysis has been issued~\cite{ecker2} which is more suited 
to our purpose, in the sense that it applies 
to the inclusive radiative rate,
$\tau\rightarrow\nu_\tau\,\pi^-\pi^0~(\gamma)$, as measured by the
experiments. A consistent calculation of radiative corrections is
presented including real photon emission and the effect of virtual loops. 
All the contributions listed in the previous section are included 
and the isospin-breaking contributions in the pion form factor are 
now more complete. Following Ref.~\cite{ecker2}, the 
relation between the Born level \ee\ \sf\   and 
the $\tau$ \sf\ (\ref{eq_v-}) reads
\beq
v_{\pi^+\pi^-}(s) \, = \, \frac{1}{G_{\rm EM}(s)}
		\,\frac{\beta_0^3}{\beta_-^3}
               \,\left|\frac{F_\pi^0(s)}{F_\pi^-(s)}
		\right|^2 v_{\pi^-\pi^0(\gamma)}(s)~,
\eeq
where $G_{\rm EM}(s)$ is the long-distance radiative correction 
involving both real photon emission and virtual loops 
(the infrared divergence cancels in the sum). Note that the 
short-distance $S_{\rm EW}$ correction, discussed above,
is already applied in the definition of $v_-(s)$ 
({\it cf.} Eq.~(\ref{eq_sf})), but its value differs from Eq.~(\ref{sew})
because subleading quark-level and hadron-level contributions should not
be added, as double counting would occur. The correct expression for the
$\pi \pi^0$ mode therefore reads
\beq
\label{sew_rho}
  S_{\mathrm{EW}}^{\pi \pi^0}(s)=\frac{S^{\rm had}_{\mathrm{EW}}G_{\rm EM}(s)}
           {S^{\rm sub,lep}_{\mathrm{EW}}}
                    =(1.0233\pm0.0006)\cdot G_{\rm EM}(s)~,
\eeq
the subleading hadronic corrections being now incorporated in the
mass-dependent $G_{\rm EM}(s)$ factor.
The form factor correction is dominated by the effect of the pion mass
difference in the $\rho$ width, but it also includes a small contribution
at the $10^{-3}$ level from the 'chiral' form used for the $\rho$ 
lineshape. In practice, however, the correction is independent of
the chosen parametrization of the form factor.
The different contributions to the isospin-breaking corrections
are shown in the second column of Table~\ref{tab_isobreak}.
The values slightly differ from those given in Ref.~\cite{ecker2}
because the authors use a model for the pion form factor
rather than integrating experimental data. The largest difference
however stems from our re-evaluation of the short-distance electroweak 
correction, $S_{\rm EW}$,  including the subleading leptonic contribution. 
The sum amounts to $(-13.8\pm2.4)~10^{-10}$ to be compared with 
$(-12.0\pm2.6)~10^{-10}$ given in Ref.~\cite{ecker2}.
\vs
The dominant uncertainty in this method stems from the $\rho^\pm$-$\rho^0$
mass difference. Indeed, in the chiral model used in Ref.~\cite{ecker2}
the only parameter entering the pion form factor is the $\rho$ mass,
since the width is given by 
$\Gamma_\rho(s)=m_\rho s\,\beta^3(s)/(96\pi f_\pi^2)$. In the method
previously used and recalled in Section~\ref{sec_isobreak1},
the width at the pole was taken as an independent parameter with 
$\Gamma_\rho(s)=\Gamma_\rho\sqrt{s}\,\beta^3(s)/(m_\rho\beta^3(m_\rho^2))$,
so that the effect of the $\rho$ mass difference approximately cancels 
after integration. This explains the large difference in the 
uncertainties quoted for the two evaluations in Table~\ref{tab_isobreak}.
\vs
Since the integral~(\ref{eq_int_amu}) requires as input the \ee\ \sf\ 
including FSR photon emission, a final correction is necessary. 
It is identical to that applied in the CMD-2 analysis~\cite{cmd2,jeger_rad}
(\cf\   Section~\ref{sec_rad}).
All the corrections are drawn versus $s$ in Fig.~\ref{fig_isobreak}. The
overall correction reduces the $\tau$ rate below the $\rho$ peak, but,
somewhat unexpectedly, has the opposite effect above. This behavior is
driven by the long-distance radiative corrections contained in 
$G_{\rm EM}(s)$.
\vs
The total correction to the $\tau$ result in this method, 
not including the FSR contribution, amounts to 
$\Delta a_\mu^{\rm had}=(-13.8 \pm 2.4)~10^{-10}$, where the main
contribution to the error is due to the experimental limits on the
$\rho$ mass difference. After
including the FSR contribution, it becomes $(-9.3 \pm 2.4)~10^{-10}$,
a value consistent with the result in the first column of 
Table~\ref{tab_isobreak} which
does not include the virtual corrections and uses a less sophisticated
treatment of radiative decays. In the following we apply the correction
functions from the more complete analysis 
(method (II) in Table~\ref{tab_isobreak}) and keep the corresponding
uncertainty separate from the purely experimental errors.

\subsection{\it Isospin Breaking in $4\pi$ Channels}
\label{sec_iso4pi}

There exists no comparable study of isospin breaking in the $4\pi$
channels. Only kinematic corrections resulting from the pion mass difference
have been considered so far~\cite{czyz}, which we have applied in
this analysis. It creates shifts of $-0.7~10^{-10}$ ($-3.8\%$) and
$+0.1~10^{-10}$ ($+1.1\%$) for $2\pi^+2\pi^-$ and $\pi^+\pi^-2\pi^0$,
respectively. However, since the four-pion contribution to \amuhadLO\ is 
relatively less important than the two-pion part 
(by a little more than an order of magnitude in the integration 
range up to 1.8~GeV) and the experimental uncertainties 
are much larger, we feel this is a justified procedure
at the present level of accuracy of the data.
Moreover, the entire correction has been attributed as systematic error
which is kept separate from the experimental errors on \amuhadLO\ from
these channels. 

It should also be pointed out that the systematic uncertainties from
isospin breaking are essentially uncorrelated between the $2\pi$ and
$4\pi$ modes: as Table~\ref{tab_isobreak} shows, the dominant sources 
of uncertainties are the $\rho^\pm$-$\rho^0$ mass difference for $2\pi$
and the threshold factors in $4\pi$ where large errors have been given
to cover uncertainties in the decay dynamics and the missing pieces.

%
%
\section{Comparison of \ee\ and $\tau$ Spectral Functions}

The \ee\ and the isospin-breaking corrected $\tau$ \sfs\ can be 
directly compared for the dominant $2\pi$ and $4\pi$ final states. For
the $2\pi$ channel, the $\rho$-dominated form factor falls off very
rapidly at high energy so that the comparison can be performed 
in practice over the full energy range of interest. The situation
is different for the $4\pi$ channels where the $\tau$ decay kinematics
limits the exercise to energies less than $\sim$~1.6~GeV, with only
limited statistics beyond.

\subsection{\it Direct Comparison}
\label{sec_compsf}

\begin{figure}[p]
\epsfxsize16cm
\centerline{\epsffile{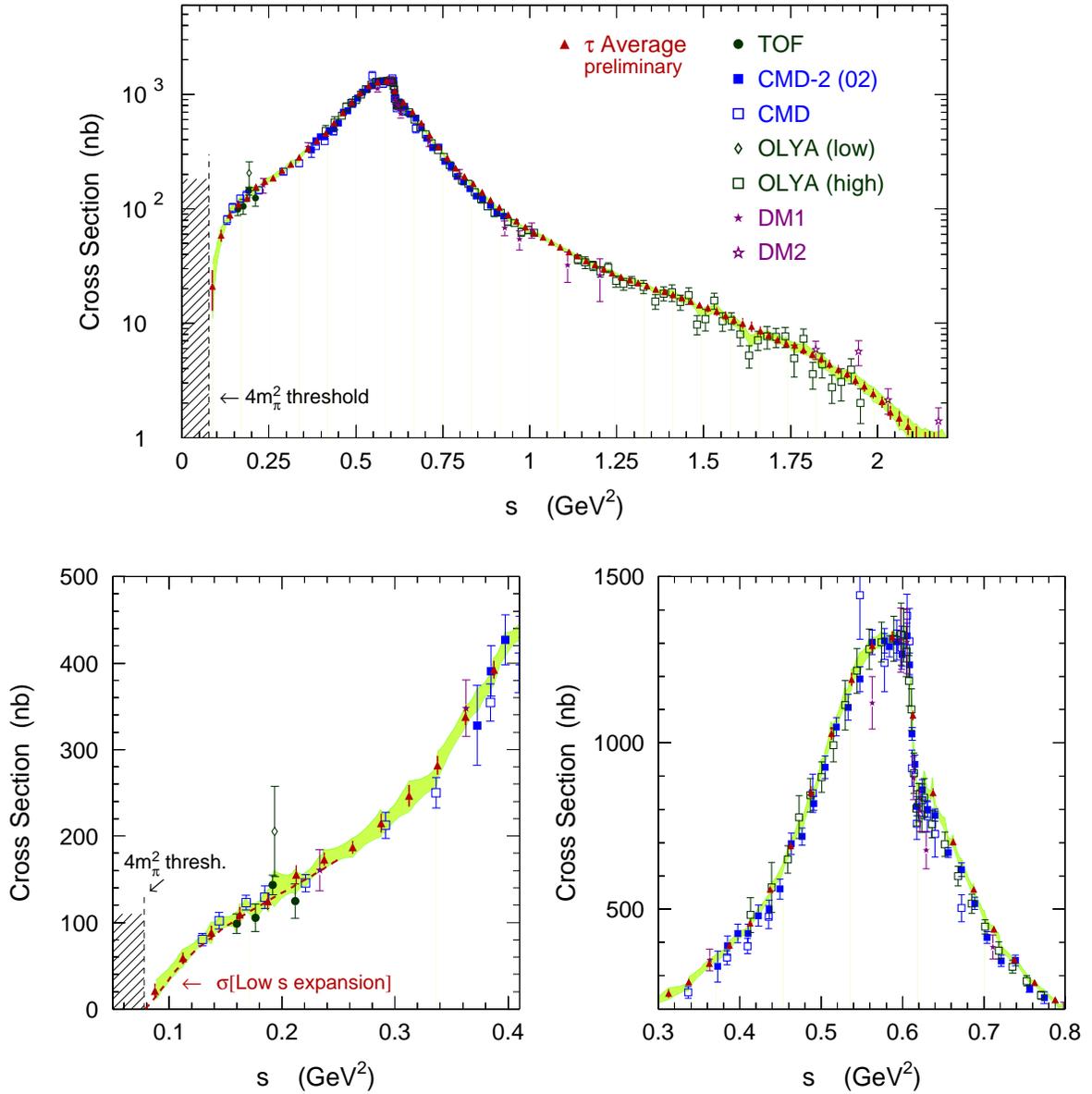}}
\caption[.]{\it Comparison of the $\pi^+\pi^-$ \sfs\
    	from \ee\  and isospin-breaking corrected $\tau$ data, 
	expressed as \ee\ cross sections. The band indicates the 
	combined \ee\ and $\tau$ result within $1\sigma$ errors.
	It is given for illustration purpose only. }
\label{fig_2pi_eetau}
\end{figure}
\begin{figure}[p]
\epsfxsize12cm
\centerline{\epsffile{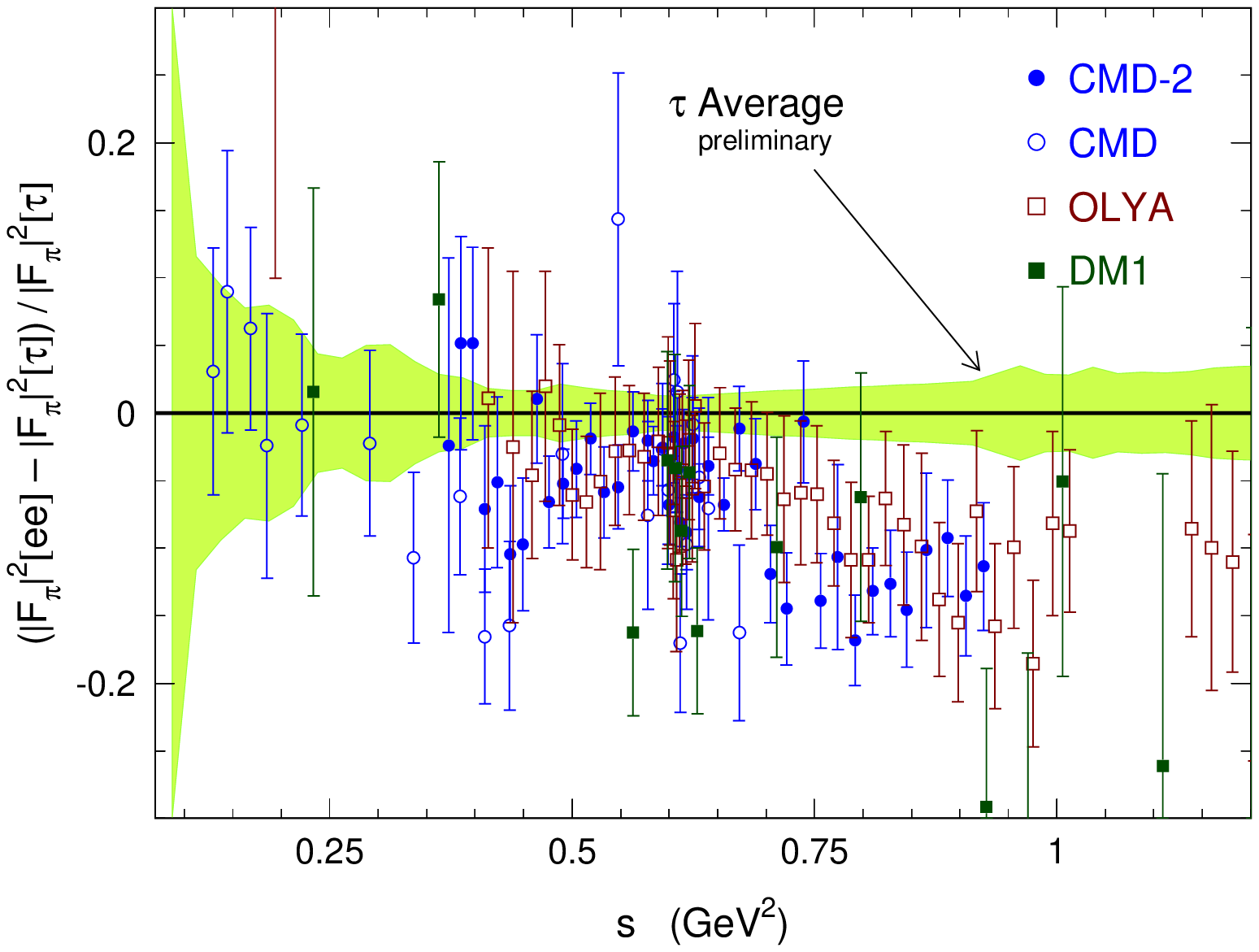}}
\caption[.]{\it Relative comparison of the $\pi^+\pi^-$ \sfs\
    	from \ee\  and isospin-breaking corrected $\tau$ data, 
	expressed as a ratio to the $\tau$ \sf.
	The band shows the uncertainty on the latter.}
\label{fig_2pi_comp}
\vspace{0.6cm}
\epsfxsize12cm
\centerline{\epsffile{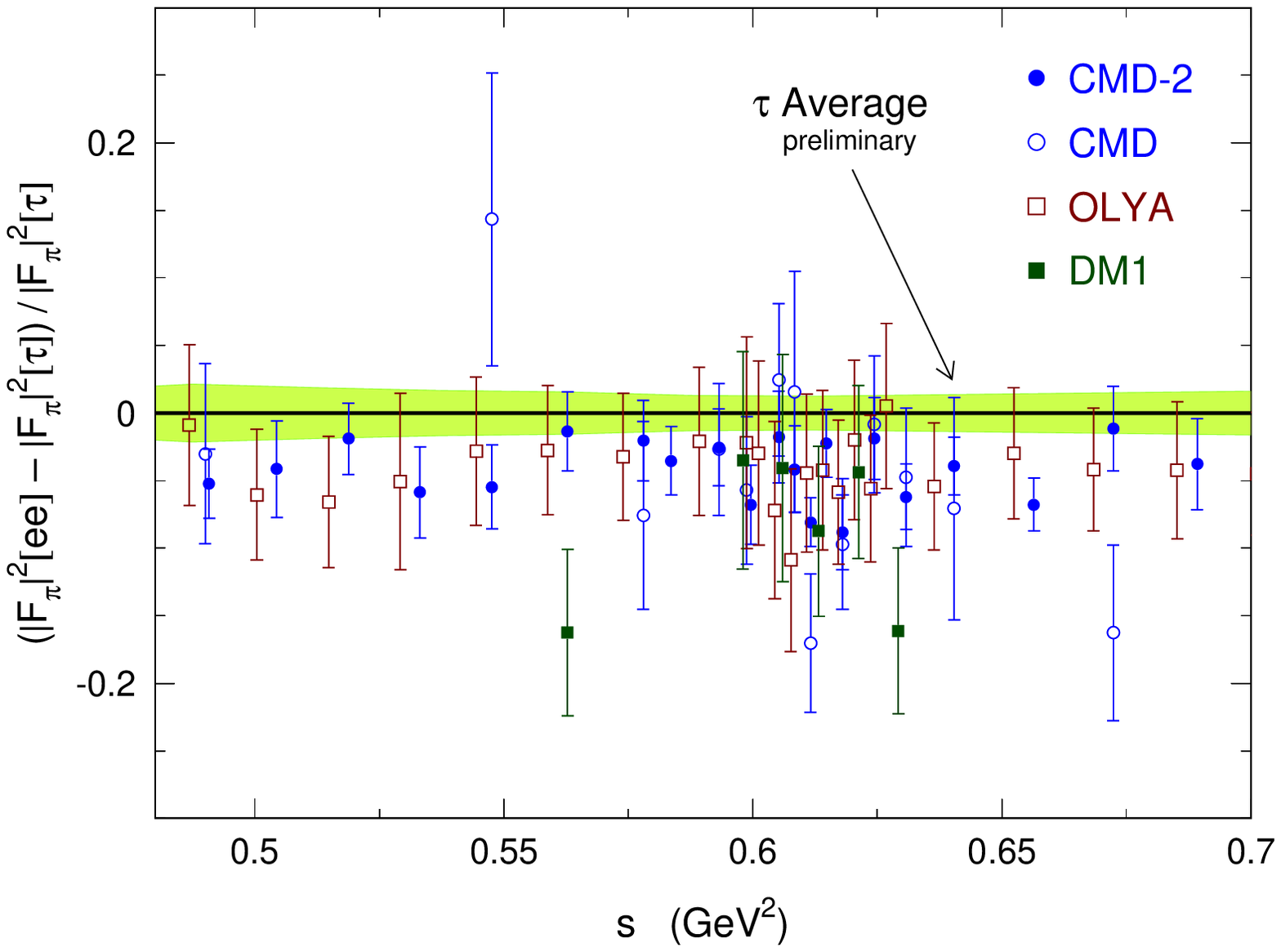}}
\caption[.]{\it Relative comparison in the $\rho$ region of the 
	$\pi^+\pi^-$ \sfs\
    	from \ee\  and isospin-breaking corrected $\tau$ data, 
	expressed as a ratio to the
     	$\tau$ \sf. The band shows the uncertainty on the latter.}
\label{fig_2pi_comp_zoom}
\end{figure}
\begin{figure}[p]
\epsfxsize12cm
\vspace{-2.3cm}
\centerline{\epsffile{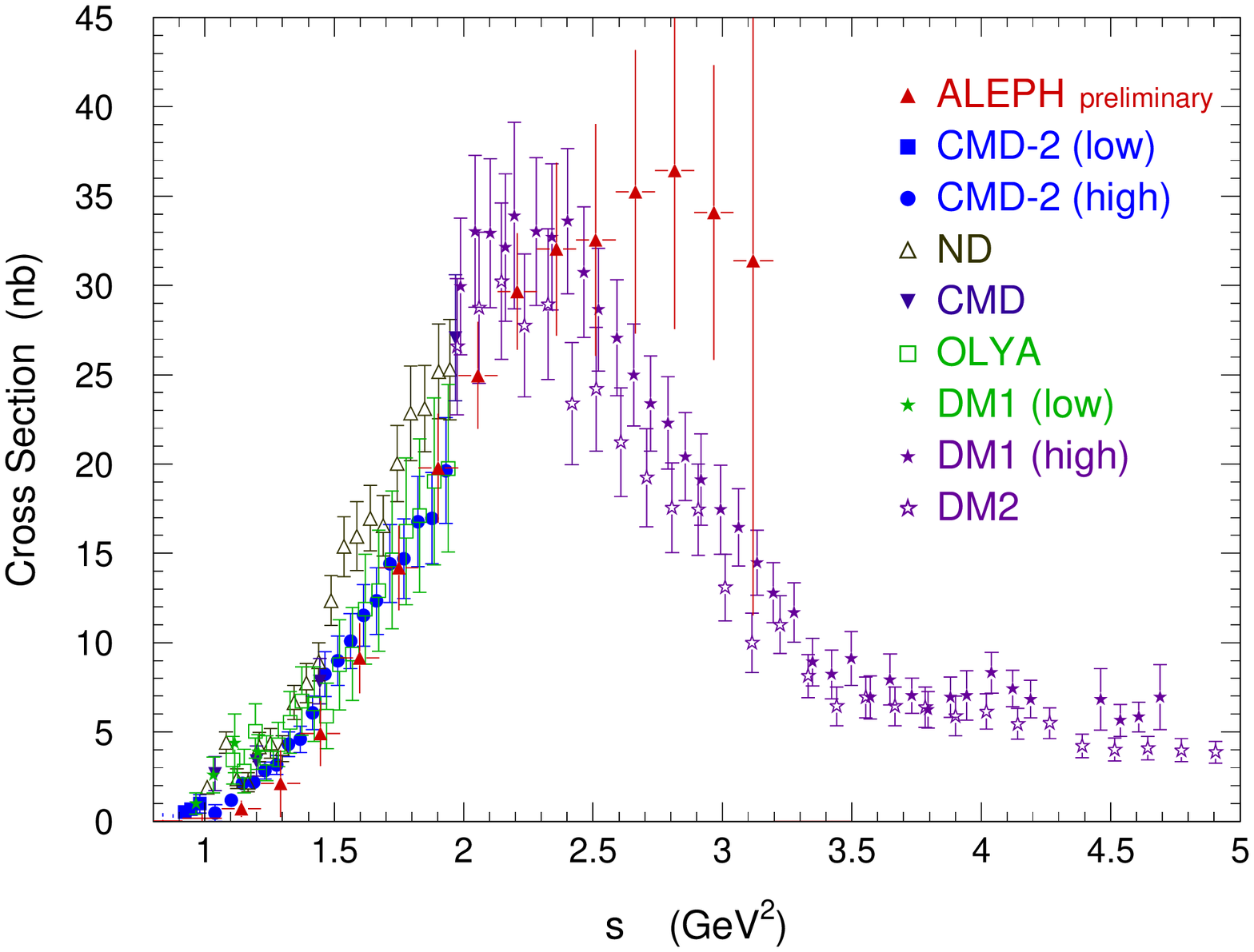}}
\vspace{-0.5cm}
\caption[.]{\it \it Comparison of the $2\pi^+2\pi^-$ \sfs\
    	from \ee\ and isospin-breaking corrected $\tau$ data,
	expressed as \ee\ cross sections.}
\label{fig_4pi_eetau}
\epsfxsize12cm
\vspace{-1.3cm}
\centerline{\epsffile{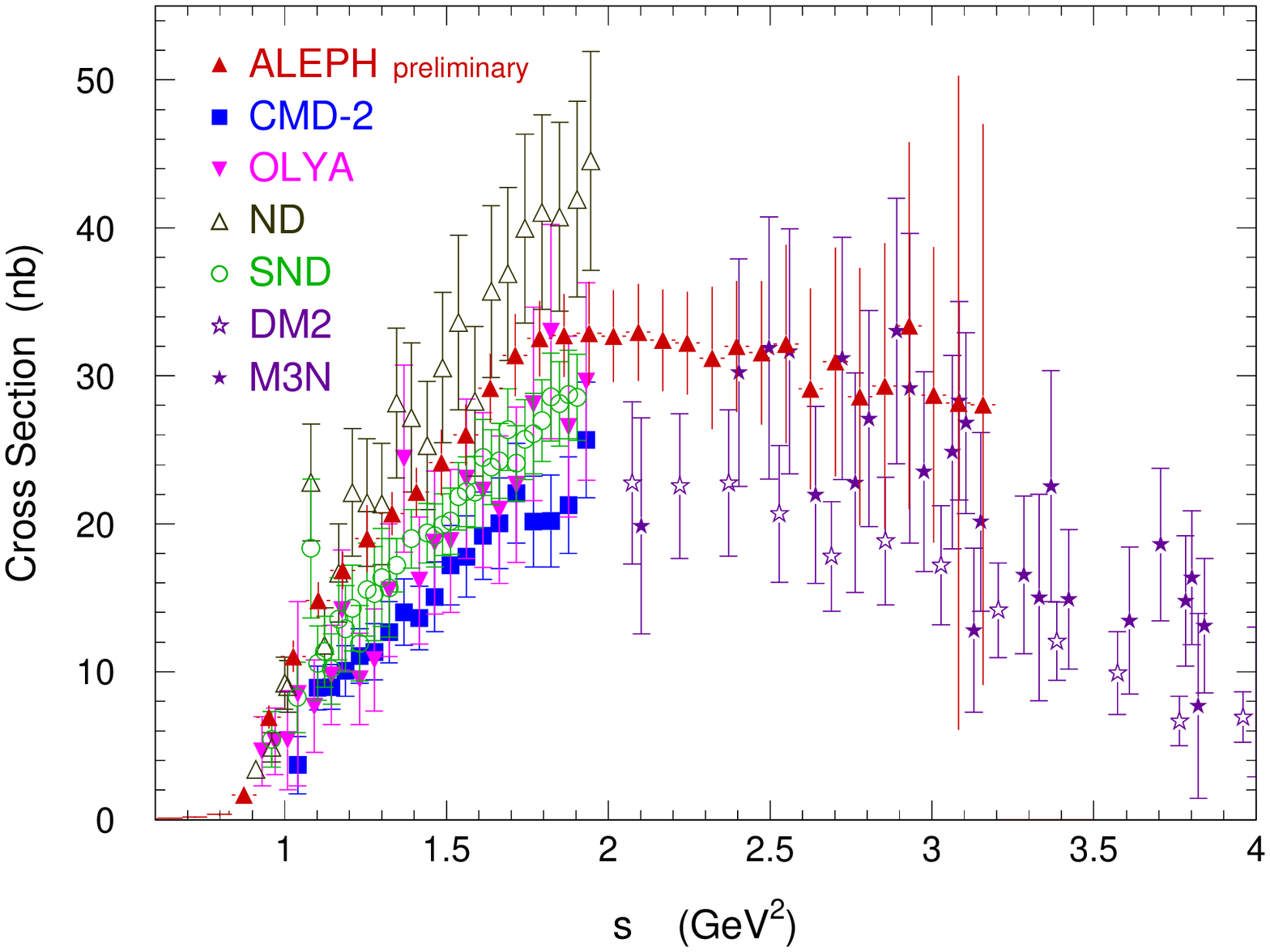}}
\vspace{-0.5cm}
\caption[.]{\it \it Comparison of the $\pi^+\pi^-2\pi^0$ \sfs\
    	from \ee\ and isospin-breaking corrected $\tau$ data,
	expressed as \ee\ cross sections.}
\label{fig_2pi2pi0_eetau}
\end{figure}

Fig.~\ref{fig_2pi_eetau} shows the comparison for the $2\pi$ \sfs. 
Visually, the agreement seems satisfactory, however the large 
dynamical range involved
does not permit an accurate test. To do so, the \ee\ data are plotted
as a point-by-point ratio to the $\tau$ \sf\ in Fig.~\ref{fig_2pi_comp}, 
and enlarged in Fig.~\ref{fig_2pi_comp_zoom}, to better emphasize the 
region of the $\rho$ peak\footnote
{
   	The central bands in Figs.~\ref{fig_2pi_comp} and
   	\ref{fig_2pi_comp_zoom} give the quadratic sum of the 
	statistical and systematic errors of the combined 
	$\tau$ spectral functions. Local bumps in these bands
	stem from increased errors when combining different 
	experiments having local inconsistencies. We use the
	procedure described in Section~\ref{sec:averagingExperiments}.
}. 
The \ee\ data are significantly lower by 2-3\% 
below the peak, the discrepancy increasing to about 10\% in the 
0.9-1.0~GeV region.
\vs
The comparison for the $4\pi$ cross sections is given 
in Fig.~\ref{fig_4pi_eetau}
for the $2\pi^+2\pi^-$ channel and in Fig.~\ref{fig_2pi2pi0_eetau} for 
$\pi^+\pi^-2\pi^0$. As noted before, the latter suffers from
large differences between the results from the different \ee\ 
experiments. The $\tau$ data, combining two measured \sfs\ according
to Eq.~(\ref{eq_cvc_2pi2pi0}) and corrected for isospin breaking as
discussed in Section~\ref{sec_cvc}, lie somewhat in between with large
uncertainties above 1.4~GeV because of the lack of statistics and a large 
feedthrough background in the $\tau\rightarrow\nu_\tau\,\pi^-3\pi^0$
mode. In spite of these difficulties the  $\pi^-3\pi^0$ \sf\ is in agreement
with \ee\ data as can be seen in Fig.~\ref{fig_4pi_eetau}. It is clear
that intrinsic discrepancies exist among the \ee\ experiments and that a
quantitative test of CVC in the $\pi^+\pi^-2\pi^0$ channel is
premature.

\subsection{\it Branching Ratios in $\tau$ Decays and CVC}
\label{sec_brcvc}

A convenient way to assess the compatibility between \ee\ and $\tau$
\sfs\ proceeds with the evaluation of $\tau$ decay fractions using
the relevant \ee\ \sfs\ as input. All the isospin-breaking corrections 
detailed in Section~\ref{sec_isobreak2} are included. The advantage of this
procedure is to allow a quantitative comparison using a single number.
The weighting of the \sf\ is however different from the vacuum
polarization kernels. Using the branching fraction
$B(\tau^-\rightarrow \nu_\tau\,e^-\,\bar{\nu}_e)\,=\,(17.810 \pm 0.039)\%$,
obtained assuming leptonic universality in the charged weak 
current~\cite{aleph_new}, the results for the main channels are given
in Table~\ref{tab_brcvc}. The errors quoted for the CVC values are split
into uncertainties from ({\it i}) the experimental input
(the \ee\ annihilation cross sections) and the numerical integration procedure,
({\it ii}) the missing radiative corrections applied to the relevant \ee\ data,
and ({\it iii}) the isospin-breaking corrections when relating $\tau$ and \ee\
\sfs. The values for the $\tau$ branching ratios involve 
measurements~\cite{aleph_new,cleo_bpipi0,opal_bpipi0}
given without charged hadron identification, {\it i.e.} for the
$h\pi^0\nu_\tau$, $h3\pi^0\nu_\tau$ and $3h\pi^0\nu_\tau$ final states. 
The corresponding channels with charged kaons have been 
measured~\cite{aleph_ksum,cleo_kpi0} and their
contributions can be subtracted out in order to obtain the pure pionic
modes. 
\vs
\begin{table}[t]
\begin{center}
\setlength{\tabcolsep}{0.75pc}
{\small
\begin{tabular}{lrrr} \hline 
&&& \\[-0.3cm]
		& \mc{3}{c}{Branching fractions  (in \%)} \\
\rs{~~~~~Mode} 	& \mc{1}{c}{$\tau$ data} 	
		& \mc{1}{c}{$e^+e^-$ via CVC} & $\Delta(\tau-e^+e^-)$ 
\\[0.15cm]
\hline
&&& \\[-0.3cm]
\mc{1}{l}{$~\tau^-\rar\nu_\tau\pi^-\pi^0$}
		& $25.46 \pm 0.12$ 
		& $23.98 \pm \underbrace{0.25_{\rm exp}	
			\pm 0.11_{\rm rad}\pm 0.12_{\rm SU(2)}}_{0.30}$ 
		& $+1.48 \pm 0.32$ 
	\\[0.7cm]
\mc{1}{l}{$~\tau^-\rar\nu_\tau\pi^-3\pi^0$}
		& $ 1.01 \pm 0.08$ 
		& $ 1.09 \pm \underbrace{0.06_{\rm exp}
		        \pm 0.02_{\rm rad}\pm 0.05_{\rm SU(2)}}_{0.08}$ 
		& $-0.08 \pm 0.11$ 
	\\[0.7cm]
\mc{1}{l}{$~\tau^-\rar\nu_\tau2\pi^-\pi^+\pi^0$}
		& $ 4.54 \pm 0.13$ 
		& $ 3.63 \pm \underbrace{0.19_{\rm exp}
			\pm 0.04_{\rm rad}\pm 0.09_{\rm SU(2)}}_{0.21}$ 
		& $+0.91 \pm 0.25$ 
	\\[0.7cm]
 \hline
\end{tabular}
}
\end{center}
\caption{\label{tab_brcvc}
	\it Branching fractions of $\tau$ vector decays into 	
	2 and 4 pions in the final state. Second column: world 
	average. Third column: inferred from \ee\ spectral
	functions using the isospin 
	relations~(\ref{eq_cvc_2pi}-\ref{eq_cvc_2pi2pi0})
        and correcting for isospin breaking. The experimental error
	of the $\pi^+\pi^-$ CVC value contains an absolute 
	procedural integration error of $\,0.08\%$.
	Experimental errors, including uncertainties on the integration 
        procedure, and theoretical (missing radiative corrections for \ee,
	and isospin-breaking corrections and $V_{ud}$ for $\tau$)
	are shown separately.
	Right column: differences between the
	direct measurements in $\tau$ decays and the CVC evaluations,
        where the separate errors have been added in quadrature.}
\end{table} 
\begin{figure}[t]
\epsfxsize12cm
\centerline{\epsffile{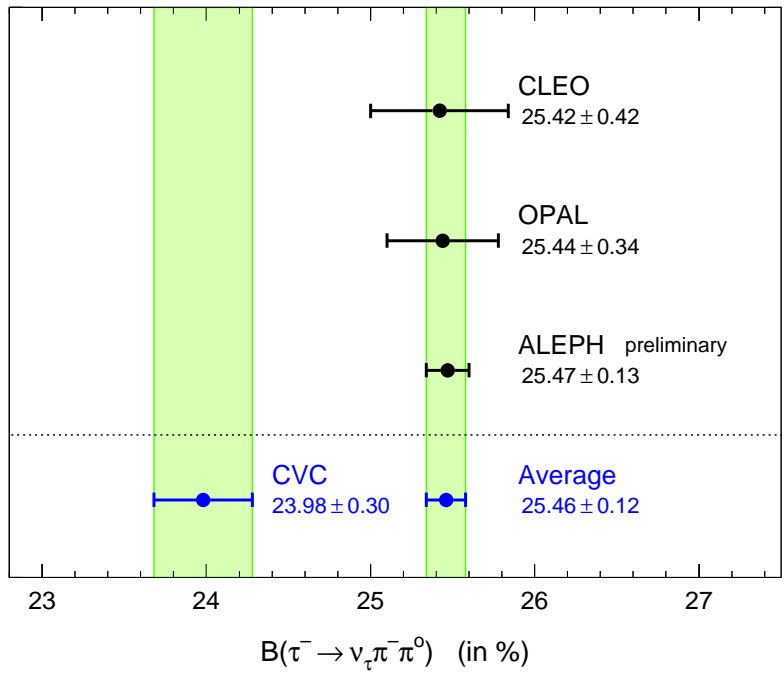}}
\caption[.]{\it The measured branching ratios for 
       $\tau\rightarrow\nu_\tau\pi^-\pi^0$ compared to the prediction
       from the $e^+e^-\rar\pi^+\pi^-$ \sf\ applying the isospin-breaking
       correction factors discussed in Section~\ref{sec_isobreak2}.
       The measured branching ratios are from ALEPH~\cite{aleph_new},
       CLEO~\cite{cleo_bpipi0} and OPAL~\cite{opal_bpipi0}.
       The OPAL result was obtained from their $h \pi^0$ branching 
       ratio, reduced by the small $K \pi^0$ contribution 
       measured by ALEPH~\cite{aleph_ksum} and CLEO~\cite{cleo_kpi0}.}
\label{fig_cvc_2pi}
\end{figure}
As expected from the preceding discussion, a large discrepancy is 
observed for the $\tau\rightarrow \nu_\tau\,\pi^-\pi^0$ 
branching ratio, with a difference of 
$(-1.48\pm0.12_\tau\pm0.25_{\rm ee}\pm0.11_{\rm rad}
\pm0.12_{\rm SU(2)})\%$, 
where the uncertainties are from the $\tau$ branching ratio, 
\ee\ cross sections, \ee\ missing radiative corrections and isospin-breaking 
corrections (including the uncertainty on $V_{ud}$), respectively. 
Adding all errors in quadrature, the effect represents a 4.6~$\sigma$ 
discrepancy. Since the disagreement between \ee\ and $\tau$ \sfs\ is 
more pronounced at energies above 750~MeV, we expect a smaller discrepancy
in the calculation of \amuhadLO\ because of the steeply falling kernel
$K(s)$ in this case. More information on the comparison is
displayed in Fig.~\ref{fig_cvc_2pi} where it is clear that ALEPH, CLEO 
and OPAL all separately, but with different significance,  
disagree with the \ee-based CVC result.
\vs
The situation in the $4\pi$ channels is different. Agreement is
observed for the $\pi^-3\pi^0$ mode within an accuracy of $11\%$, however
the comparison is not satisfactory for the $2\pi^-\pi^+\pi^0$ mode. 
In the latter case, the relative difference is very large, 
$(22\pm6)$\%, compared to any reasonable level of isospin symmetry 
breaking. As such, it rather points to experimental problems that have
to be investigated.

%
%

\section{The Integration Procedure}
\label{sec_integration}

The information used for the evaluation of the integral~(\ref{eq_int_amu}) 
comes mainly from direct measurements of the cross sections in \ee\ 
annihilation and \via\ CVC from $\tau$ \sfs. In general, the integrals 
themselves are evaluated using the trapezoidal rule, {\it i.e.}, 
combining adjacent measurement points by linear interpolation. Even if this 
method is straightforward and free from theoretical assumptions (other 
than CVC in the $\tau$ case), its numerical calculation requires special 
care. The finite and variable distance between adjacent measurements
creates systematic uncertainties that have to be estimated.
The combination of measurements from different experiments taking
into account correlations---both within each data set and between
different experiments---is the subject of additional discussions 
presented in the following.

%
%
\subsection{\it Averaging Data from Different Experiments}
\label{sec:averagingExperiments}

To exploit the maximum information from the available data, we combine 
weighted measurements of different experiments at a given energy instead 
of calculating separately the integrals for every experiment and finally 
averaging them.
\vs
The solution of the averaging problem is found by minimizing
\beq\label{eq_chi2}
     \chi^2 \:=\: \sum_{n=1}^{N_{\mathrm exp}} 
                      \sum_{i,j=1}^{N_n}
                          (x_i^n-k_i)\,(C_{ij}^n)^{-1}\,(x_j^n-k_j)~,
\eeq
where $x_i^n$ is the $i$th cross section measurement of the $n$th experiment
in a given final state, $C_{ij}^n$ is the covariance between the $i$th and 
the $j$th measurement and $k_i$ is the unknown distribution to be determined. 
The covariance matrix $C^n$ is given by
\beq\label{eq_cov}
     C_{ij}^n \:=\: \Bigg\{ 
                    \begin{array}{l@{\quad\quad}l}
                      (\Delta_{i,\mathrm stat}^n )^2 +
                        (\Delta_{i,\mathrm sys}^n )^2 ~~~{\mathrm for}~i=j \\
                      \Delta_{i,\mathrm sys}^n \cdot
              \Delta_{j,\mathrm sys}^n  ~~~~~~\,{\mathrm for}~i\neq j
                    \end{array} ,~~~i,j=1,\dots,N_n~,
\eeq
where $\Delta_{i,\mathrm stat}^n$ 
($\Delta_{i,\mathrm sys}^n$) denotes the statistical (systematic) error
of $x_i^n$. The systematic errors of the \ee\ annihilation measurements 
are essentially due to luminosity and efficiency uncertainties.
It is conservative to take them as common errors of all data points
of a given experiment.
The minimum condition $d\chi^2/dk_i=0,~\forall i$ leads to the system 
of linear equations
\beq\label{eq_dchi2}
   \sum_{n=1}^{N_{\mathrm exp}} 
                      \sum_{j=1}^{N_n}\,
                          (x_j^n-k_j)\,(C_{ij}^n)^{-1} 
                          \:=\: 0~,~~~~~i=1,\dots,N_n~.
\eeq
The inverse covariance $\tilde{C}_{ij}^{-1}$ between the solutions $k_i$, 
$k_j$ is the sum of the inverse covariances of each experiment
\beq\label{eq_ddchi2}
    \tilde{C}_{ij}^{-1} \:=\: \sum_{n=1}^{N_{\mathrm exp}} (C_{ij}^n)^{-1}~.
\eeq
If different measurements at a given energy show 
inconsistencies, {\it i.e.}, their $\chi^2$ per number of degrees of 
freedom (DF) is larger than one, we rescale the error of their 
weighted average by $\sqrt{\chi^2/\mathrm DF}$.
 
%
%
\subsection{\it Correlations between Experiments}

Eq.~(\ref{eq_ddchi2}) provides the covariance matrix needed for the 
error propagation when calculating the integrals over the solutions 
$k_i$ from Eq.~(\ref{eq_dchi2}). Up to this point, $\tilde{C}_{ij}$
only contains correlations between the systematic uncertainties within
the same experiment. However, due to commonly used simulation techniques
for the acceptance and luminosity determination as well as state-of-the-art
calculations of radiative corrections, systematic correlations from 
one experiment to another occur. It is obviously a 
difficult task to reasonably estimate the amount of such correlations
as they depend on the reconstruction capabilities of the experiments and
the theoretical understanding of the underlying decay dynamics. 
In general, one can state that in older
experiments, where only parts of the total solid angle were covered
by the detector acceptance, individual experimental limitations should
dominate the systematic uncertainties. Potentially common systematics, such
as radiative corrections or efficiency, acceptance and luminosity
calculations based on the Monte Carlo simulation, play only minor roles. 
The correlations between systematic errors below 2~GeV energy are 
therefore estimated to be between 10\pc\ and 30\pc, with the exception of 
the $\pi^+\pi^-$ final state, where we impose a 40\pc\ correlation 
due to the simpler experimental situation and the better knowledge of 
the dynamics which leads to non-negligible systematic contributions from 
the uncertainties of the radiative corrections. At energies above 2~GeV 
the experiments measured the total inclusive cross section ratio $R$. 
Between 2 and 3~GeV, individual technical problems dominate 
the systematic uncertainties. At higher energies, new experiments provide 
nearly full geometrical acceptance which decreases the uncertainty of 
efficiency estimations. Radiative corrections as well as theoretical
errors of the luminosity determination give important contributions
to the final systematic errors quoted by the experiments. We therefore
estimate the correlations between the systematic errors of the experiments 
to be negligible between 2~GeV and 3~GeV, 20\pc\ between 3~GeV and
10~GeV. 
These correlation coefficients are added to all those entries of 
$\tilde{C}_{ij}$ from Eq.~(\ref{eq_ddchi2}) which involve two different 
experiments.

%
%
\subsection{\it Evaluation of the Integral}

The procedure described above provides the weighted average and the covariance 
of the cross sections from different experiments contributing to a 
certain final state in a given range of energies. We now apply the 
trapezoidal rule. To perform the 
integration~(\ref{eq_int_amu}), we subdivide the integration range in 
fine energy steps and calculate for each of these steps 
the corresponding covariance (where additional correlations induced by 
the trapezoidal rule have to be taken into account). This procedure yields
error envelopes between adjacent measurements as depicted by the shaded
bands in the corresponding figures.
\vs
As a cross check, a different procedure of the evaluation of the 
integral has been applied. For each final state, results of different 
experiments contributing to it in a given energy range are
integrated separately using a rectangular method. After that a weighted
average, based on the statistical and systematic errors combined 
in quadrature, is calculated. In some cases when correlations between
systematic uncertainties of different experiments are known, they
are taken into account after averaging the results with weights based 
on the statistical errors only. As mentioned above, if results of 
the integration for different measurements are found to be inconsistent, 
the error is rescaled by a factor $\sqrt{\chi^2/{\rm DF}}$. 
\vs
The difference between the results of the
two described procedures is considered when estimating the 
systematic uncertainty on the numerical integration procedure. 
The systematics also take into account variations of the energy 
interval where several data points are lumped into a single value,
and the effect on the central value of the integral when including 
or not the correlations.   
The procedural systematics are added in quadrature to the experimental 
error on the integral.
For instance, in the case of the $\pi^+\pi^-$ contribution, this
procedural uncertainty amounts to $1.5~10^{-10}$. 

%
%
\section{Specific Contributions}

In some energy regions where data information is scarce and reliable 
theoretical predictions are available, we use analytical contributions
to extend the experimental integral. Also, the treatment of narrow
resonances involves a specific procedure.
%
%
\subsection{\it The $\pi^+\pi^-$ Threshold Region}

\begin{figure}[t]
\epsfxsize12cm
\centerline{\epsffile{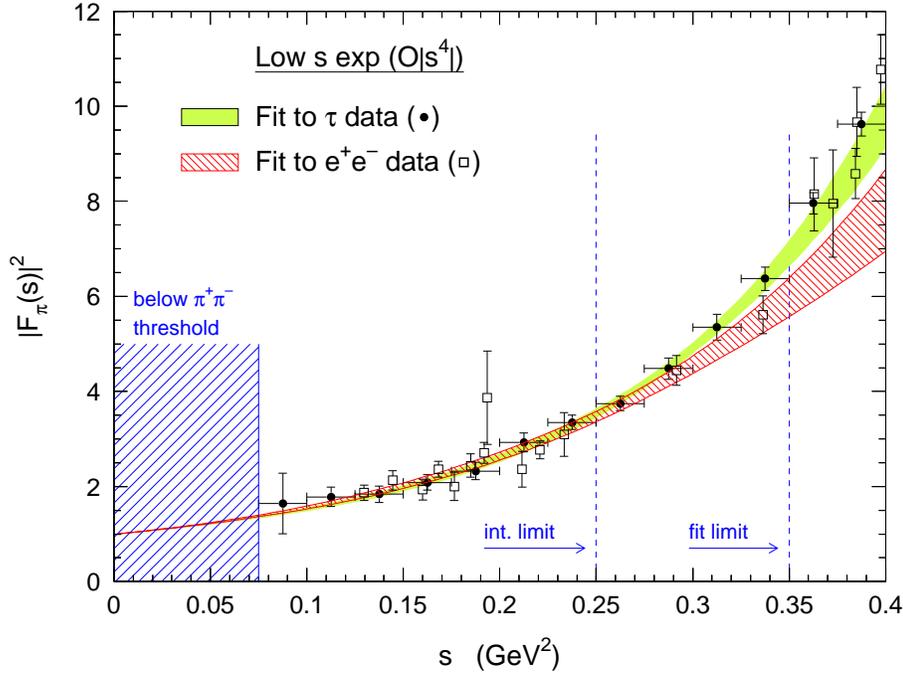}}
\caption[.]{\it Fit of the pion form factor from $4 m_\pi^2$ to
	$0.35~{\rm GeV}^2$ using a third-order Taylor expansion with 
	the constraints at $s=0$ and the measured
    	pion r.m.s. charge radius from space-like data~\cite{space_like}.
	The result of the fit is integrated only up to $0.25~{\rm GeV}^2$.}
\label{taylor}
\end{figure}

To overcome the lack of precise data at threshold energies and to 
benefit from the analyticity property of the pion form factor, 
a third order expansion in $s$ is used. The pion form factor $F_\pi^0$ 
is connected with the $\pi^+\pi^-$ cross section \via\ the expression
\beq 
|F^0_\pi|^2 \;=\; \frac{3s}{\pi\alpha^2 \beta_0^3}~\sigma_{\pi\pi}~.
\eeq
The expansion for small $s$ reads
\beq\label{eq_taylor}
F^0_{\pi} \;=\; 
      1 + \frac{1}{6}\langle r^2 \rangle_\pi\,s + c_1\,s^2 +c_2\,s^3 +
      O(s^4)~.
\eeq
Exploiting precise results from space-like data~\cite{space_like}, the 
pion charge radius-squared is constrained to 
$\langle r^2 \rangle_\pi=(0.439\pm0.008)~{\rm fm}^2$ and the 
two parameters $c_{1,2}$ are fitted to the data in the range 
[$2m_\pi$, 0.6~GeV]. In the case of $\tau$ data, isospin corrections 
are taken into account as discussed before.
\vs
The results of the fits are given in Table~\ref{tab_taylor} and shown 
in Fig.~\ref{taylor}. Good agreement is observed  
in the low energy region where the expansion should be reliable. Since 
the fits incorporate unquestionable constraints from first principles, 
we have chosen to use this parameterization for evaluating the integrals in
the range up to 0.5~GeV. Systematic uncertainties due to the fitting procedure
(fit boundaries, whether or not the coefficient $c_2$ is fixed) are small,
albeit taken into account.

\begin{table}[t]
\begin{center}
\setlength{\tabcolsep}{0.6pc}
{\normalsize
\begin{tabular}{ccccccc} \hline 
&&&&&& \\[-0.3cm]
&&&&&& \amuhadLO\ $(10^{-10})$ \\
\rs{Data}	& \rs{Coefficient} & \rs{Fit result} 
		& \mc{3}{c}{\rs{Correlation matrix}}
		& {\footnotesize $[2m_{\pi^\pm}-0.5~{\rm GeV}]$} \\[0.15cm]
\hline
&&&&&& \\[-0.3cm]
	& $\langle r^2 \rangle_\pi$
	& $ (0.439 \pm 0.008)~{\rm fm}^2$  & $ 1$ & $\star$ & $\star$ & \\
\ee	& $c_1$
	& $ (6.8 \pm 1.9)~{\rm GeV^{-4}}$  & $-0.15$ & $ 1$ & $\star$ 
		& $58.0 \pm 1.7 \pm 1.1_{\rm rad}$ \\
	& $c_2$
	& $ (-0.7 \pm 6.8)~{\rm GeV^{-6}}$  & $ 0.09$ & $-0.97$ & $ 1$ & \\[0.15cm]
\hline
&&&&&& \\[-0.3cm]
	& $\langle r^2 \rangle_\pi$
	& $(0.439 \pm 0.008)~{\rm fm}^2$  & $ 1$ & $\star$ & $\star$ & \\
$\tau$	& $c_1$
	& $ (3.3 \pm 1.7)~{\rm GeV^{-4}}$  & $-0.15$ & $ 1$ & $\star$ 
		& $56.0 \pm 1.6 \pm 0.3_{\rm SU(2)}$ \\
	& $c_2$
 	& $(13.2 \pm 5.7)~{\rm GeV^{-6}}$  & $ 0.09$ & $-0.99$ & $ 1$ & \\[0.15cm]
 \hline
\end{tabular}
}
\end{center}
\caption[.]{\label{tab_taylor}
	\it Fit results of the low energy 
	expansion~(\ref{eq_taylor}) 
	to \ee\ and $\tau$ data, the latter corrected for SU(2) 
	breaking. The right column quotes the contributions to
	\amuhadLO, integrated from threshold to 0.5~GeV.
	The errors are dominated by experiment, but take into 
	account systematic uncertainties from the fitting procedure
	(mainly the variation of the upper energy cut yielding, e.g.,  
	an uncertainty of about $0.49\,10^{-10}$ for $\tau$ data).
	The systematics in \amuhadLO\  
	from radiative corrections (\ee) and
	isospin breaking ($\tau$) (cf.   Sections~\ref{sec_rad},
	\ref{sec_isobreak1}) are quoted apart. }
\end{table} 

%
%
\subsection{\it Integration over the $\omega$ and $\phi$ Resonances}
\label{sec_omphi_int}

\begin{figure}[t]
\centerline{\epsfxsize7.8cm\epsffile{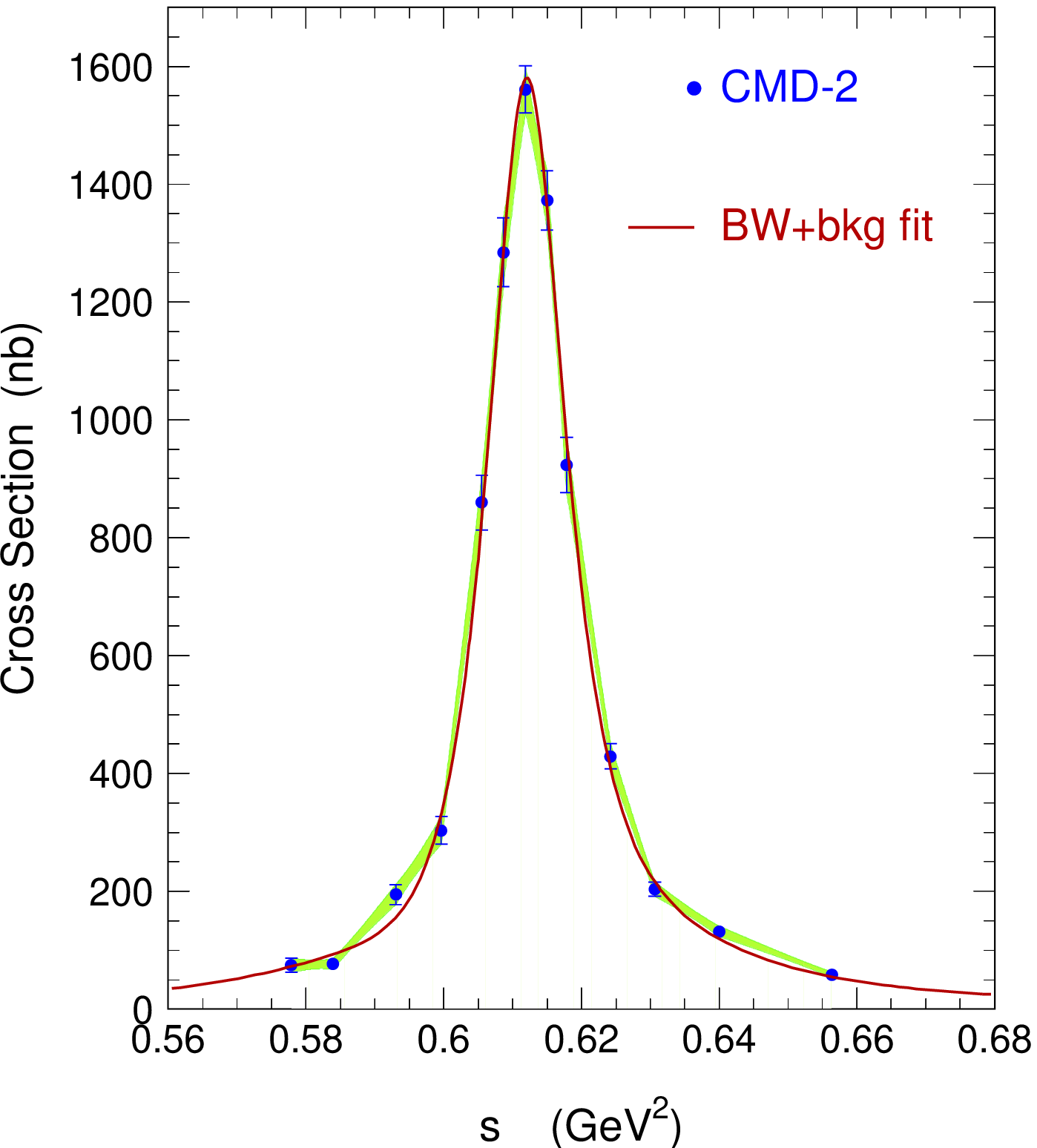}\hspace{0.2cm}
	    \epsfxsize7.8cm\epsffile{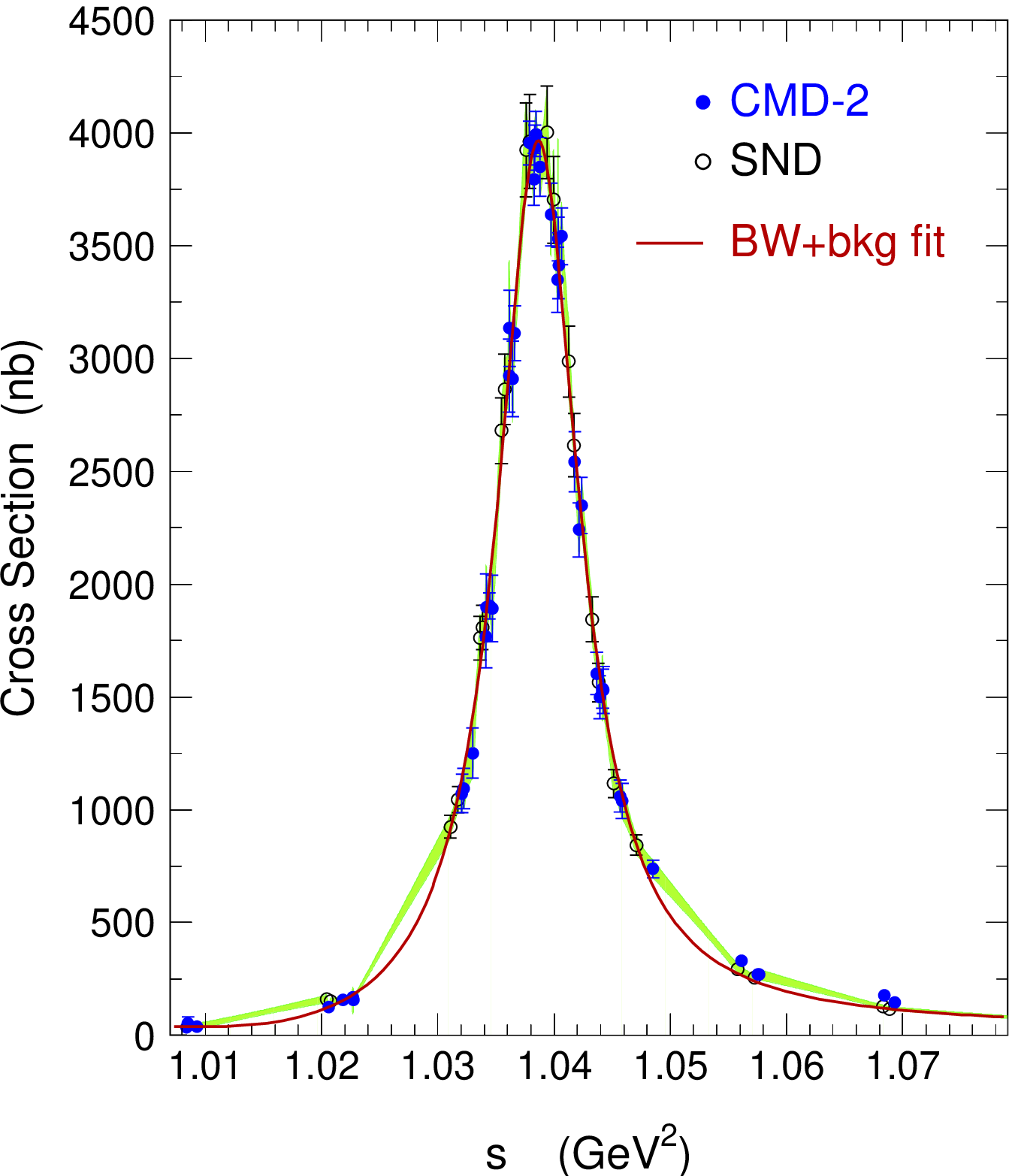}}
\caption[.]{\it Cross sections of the $\omega$ (left) and $\phi$ (right)
	resonances. The dots with error bars depict the measurements,
	the shaded band is the result of the trapezoidal rule within 
	(correlated) errors and 
	the function shows the phenomenological fit of a BW resonance 
	plus one ($\omega$) or two ($\phi$) Gaussians to account for 
	other than the single resonance contributions. The bias of the 
	trapezoidal rule when applied to a strongly concave (or convex)
	distribution is particularly visible in the tails of the $\phi$ 
        resonance when comparing to the BW fit. It leads to an overestimation
	of the integral. }
\label{fig:OmegaPhi}
\end{figure} 
\begin{table}[t]
\begin{center}
\setlength{\tabcolsep}{0.96pc}
{\small
\begin{tabular}{rrccccc} \hline
&&&&&&\\[-0.3cm]
\mc{2}{c}{\amuhadLO\ $(10^{10})$}
	& \mc{3}{c}{$\sigma($\amuhadLO$)$ $(10^{10})$} 
		& Energy range & \\
BW Fit	& Trapez.
		& Exp.	& Fit 	& BR
		& (GeV)	& \rs{Type/Ref} \\[0.15cm]
\hline
&&&&&&\\[-0.2cm]
\mc{7}{l}{\boldmath$\omega$} \\
\hline
&&&&&&\\[-0.3cm]
34.42	& 35.45	& 0.63	& 0.37	& 0.27	& 0.760184 - 0.810
	& CMD-2~\cite{cmd2_om} \\
2.51	& -	& 0.06	& 0.30	& 0.02	& 0.300 - 0.760184
	& BW fit \\[0.15cm]
\hline
&&&&&&\\[-0.3cm]
36.94	& 37.96	& \mc{3}{c}{$0.84_{\rm tot}\pm0.73_{\rm VP}\pm0.30_{\rm FSR}$}	
	& 0.300 - 0.810
	& Sum \\[0.15cm]
\hline
&&&&&&\\[-0.2cm]
\mc{7}{l}{\boldmath$\phi$} \\
\hline
&&&&&&\\[-0.3cm]
33.42	& 34.89	& 1.72	& 0.37	& 0.30	& 1.01017 - 1.03948
	& SND~\cite{snd_phi} \\
32.84	& 34.28	& 0.72	& 0.39	& 0.49	& 1.01017 - 1.03948
	& CMD-2~\cite{cmd2_phi} \\[0.15cm]
32.93	& 34.37	& \mc{3}{c}{$0.91_{\rm tot}$}
					& 1.01017 - 1.03948
	& Average \\[0.15cm]
\hline
&&&&&&\\[-0.3cm]
0.77	& -	& 0.02	& 0.07	& 0.01	& 1 - 1.01017 
	& BW fit \\
-	& 1.10	& 0.06	& 0.01	& 0.01	& 1.03948 - 1.055
	& SND~\cite{snd_phi} \\[0.15cm]
\hline
&&&&&&\\[-0.3cm]
34.80	& 36.24	& \mc{3}{c}{$0.92_{\rm tot}\pm0.63_{\rm VP}\pm0.14_{\rm FSR}$}	
	& 1 - 1.055
	& Sum \\[0.15cm]
\hline
\end{tabular}
}
\end{center}
\caption{\label{tab:OmegaPhi}\it
	Contributions to \amuhadLO\ from the narrow resonances
	$\omega(782)$ (upper table) and $\phi(1020)$ (lower table). 
	Given are the results for the BW fit (first column) and 
	the trapezoidal rule (second column). The next three columns
	quote the experimental errors, the fit parameterization
	systematics and the uncertainty introduced by the correction
	for the missing decay modes of the resonances. The energy
	interval of the integration and the integration type (data
	or analytical function) are given in the last two columns. 
	Systematic errors from the same sources, but for different
	energy regions are added linearly in the sum. All other
	errors are added in quadrature, the total errors being 
	labelled 'tot'. Additional systematics
	are due to the vacuum polarization (VP)
	correction, taken to be half of the full correction, and 
	to final state radiation (FSR) where the full correction 
	is accounted as uncertain.
}
\end{table}                                                                     

In the regions around the $\omega$ and $\phi$ resonances
we have assumed in the preceding 
works that the cross section of the $\pi^+\pi^-\pi^0$ production on the
one hand, and the $\pi^+\pi^-\pi^0$, $K^+K^-$ as well as $K^0_SK^0_L$
production on the other hand is saturated by the corresponding 
resonance production. In a data driven approach it is however more careful
to directly integrate the measurement points without introducing 
prior assumptions on the underlying process dynamics~\cite{teubner}. 
Possible non-resonant contributions and interference effects are thus 
accounted for. 
\vs
Notwithstanding,
a straightforward trapezoidal integration buries the danger of a bias:
with insufficient scan density, the linear interpolation of the 
measurements leads to a significant overestimation of the integral
when dealing with strongly concave functions such as the tails of 
Breit-Wigner resonance curves. This effect is particularly visible
in the right hand plot of Fig.~\ref{fig:OmegaPhi}, showing the
$\phi$ resonance: the cross sections are measured by 
SND~\cite{snd_phi} (sum of the final states $K^+K^-$, $K^0_SK^0_L$
and $\pi^+\pi^-\pi^0$ and corrected for missing modes, \ie,
rescaled by $(0.984\pm0.009)^{-1}$~\cite{pdg2002}) and 
CMD-2~\cite{cmd2_phi} ($K^0_SK^0_L$ only, rescaled by 
$(0.337\pm0.005)^{-1}$~\cite{pdg2002}). Shown in addition are the 
error band of the trapezoidal rule and the solution of a 
phenomenological fit of a BW resonance plus two Gaussians (only one 
Gaussian is necessary for the $\omega$,
see left hand plot in Fig.~\ref{fig:OmegaPhi}) 
to account for contributions other than the single
resonance. Both fits result in satisfactory $\chi^2$ values.
Since we are only interested in the integral and do not want to 
extract dynamical parameters like phases or branching 
fractions, it is not necessary to parametrize the exact structure
of the physical processes. We have accounted for
the systematics due to the arbitrariness in the choice of the 
parametrization by varying the functions and parameters used. The 
resulting effects are numerically small compared to the experimental
errors (see Table~\ref{tab:OmegaPhi}). It is clear from 
Fig.~\ref{fig:OmegaPhi}
that the fit function passes below the trapezoidal bands in the 
concave tails of both the $\phi$ and the $\omega$.
\vs
Table~\ref{tab:OmegaPhi} gives the contributions to \amuhadLO\
from the different energy domains covered by the experiments for 
both the $\omega$ and the $\phi$. Since the experiments quote the 
cross section results without correcting for leptonic and hadronic 
vacuum polarization in the photon propagator (\cf\   the discussion 
in Section~\ref{sec_rad}), we perform the correction
here. Note that the data shown in Fig.~\ref{fig:OmegaPhi} have been 
corrected for vacuum polarization. A small FSR correction 
(\cf\   Section~\ref{sec_rad}) is applied to the results given in 
Table~\ref{tab:OmegaPhi}.
The correction of hadronic vacuum polarization being iterative and 
thus only approximative, we assign half of the total vacuum polarization 
correction as generous systematic errors (\cf\   Section~\ref{sec_rad}). 
In spite of that, the evaluation of \amuhadLO\  is dominated by the 
experimental uncertainties. Since the trapezoidal rule is biased, we 
choose the results based on the BW fits for the final 
evaluation of \amuhadLO.

%
%
\subsection{\it Narrow $c\overline{c}$ and $b\overline{b}$ Resonances}
\label{sec_psi}

The contributions from the narrow $J/\psi$ resonances are computed
using a relativistic Breit-Wigner parametrization for their line shape.
The physical values for the resonance parameters and their errors are
taken from the latest compilation in Ref.~\cite{pdg2002}. Vacuum 
polarization effects are already included in the quoted leptonic widths. 
The total parametrization errors are then calculated by Gaussian error 
propagation. This integration procedure is not followed for the $\psi(3S)$
state which is already included in the $R$ measurements, and for the 
$\Upsilon$ resonances which are represented in an average sense (global 
quark-hadron duality) by the $b \overline{b}$ QCD contribution, discussed 
next.

%
%
\subsection{\it QCD Prediction at High Energy}
\label{sec_qcd}

Since the emphasis in this paper is on a complete and critical evaluation
of \sfs\ from low-energy data, we have adopted the conservative choice
of using the QCD prediction only above an energy of 5~GeV. The details of the 
calculation can be found in our earlier publications~\cite{dh97,dh98} and in
the references therein. Only a very brief summary shall be given here.
\vs
The perturbative QCD prediction uses a next-to-next-to-leading order
$O(\alpha_s^3)$ expansion of the Adler $D$-function~\cite{adler}, 
with second-order quark mass corrections included~\cite{kuhnmass}. 
$R(s)$ is obtained by evaluating numerically 
a contour integral in the complex $s$ plane. Nonperturbative effects
are considered through the Operator Product Expansion, giving
power corrections controlled by gluon and quark condensates. The value
$\alpha_s(M^2_{Z}) =  0.1193 \pm0.0026$, used for the evaluation 
of the perturbative part, is taken as the average of the results from
the analyses of $\tau$ decays~\cite{aleph_asf} and of the $Z$ 
width in the global electroweak fit~\cite{lepewwg}. The two determinations
have comparable uncertainties (mostly theoretical for the $\tau$ and
experimental for the $Z$) and agree well with each other. We conservatively
take as final uncertainty the value quoted in either analysis. As for the
other contributions, uncertainties are taken to be equal to half of the 
quark mass corrections and to the full nonperturbative contributions.
\vs
A test of the QCD prediction can be performed in the energy range between
1.8 and 3.7 GeV. The contribution to \amuhadLO\ in this region is computed
to be $(33.87\pm0.46)~10^{-10}$ using QCD, to be compared with the 
result, $(34.9\pm1.8)~10^{-10}$ from the data. The two values agree
within the $5\%$ accuracy of the measurements.
\vs
In Ref.~\cite{dh98} the evaluation of \amuhadLO\ was shown to be improved
by applying QCD sum rules. We do not consider this
possibility in the present analysis for the following two reasons. First,
it is clear that the main problem at energies below 2~GeV is now the
inconsistency between the \ee\ and $\tau$ input data, and this must be 
resolved with priority. Second, the improvement provided by the use of QCD sum 
rules results from a balance between the experimental accuracy of the data
and the theoretical uncertainties. The present precision of both \ee\ and
$\tau$ data, should they agree, is such that the gain would be smaller than
before. This state of affairs will be reconsidered when the problems with 
the input data are sorted out.

%
%
\section{Results}
\label{sec_results}

%
%
\subsection{\it Lowest Order Hadronic Contributions}

Before adding up all the contributions to \amuhadLO, we shall summarize
the procedure. On the one hand, the \ee-based evaluation is done in three
pieces: the sum of exclusive channels below 2~GeV, the $R$ measurements
in the 2-5~GeV range and the QCD prediction for $R$ above. Major
contributions stem from the $2\pi$ (73\%) and the two $4\pi$ ($4.5\%$) 
channels. On the other hand, in the $\tau$-based evaluation, the latter 
three contributions are taken from $\tau$ data up to 1.6~GeV and 
complemented by \ee\ data above, because the $\tau$ \sfs\ run out 
of precision near the kinematic limit of the $\tau$ mass. Thus, 
for nearly $77\%$ of \amuhadLO\  (also contributing $80\%$ of the 
total error-squared), two independent evaluations (\ee\ and $\tau$) 
are produced, the remainder being computed from \ee\ data and QCD alone.
\vs
\begin{figure}
\epsfxsize16.8cm
\centerline{\epsffile{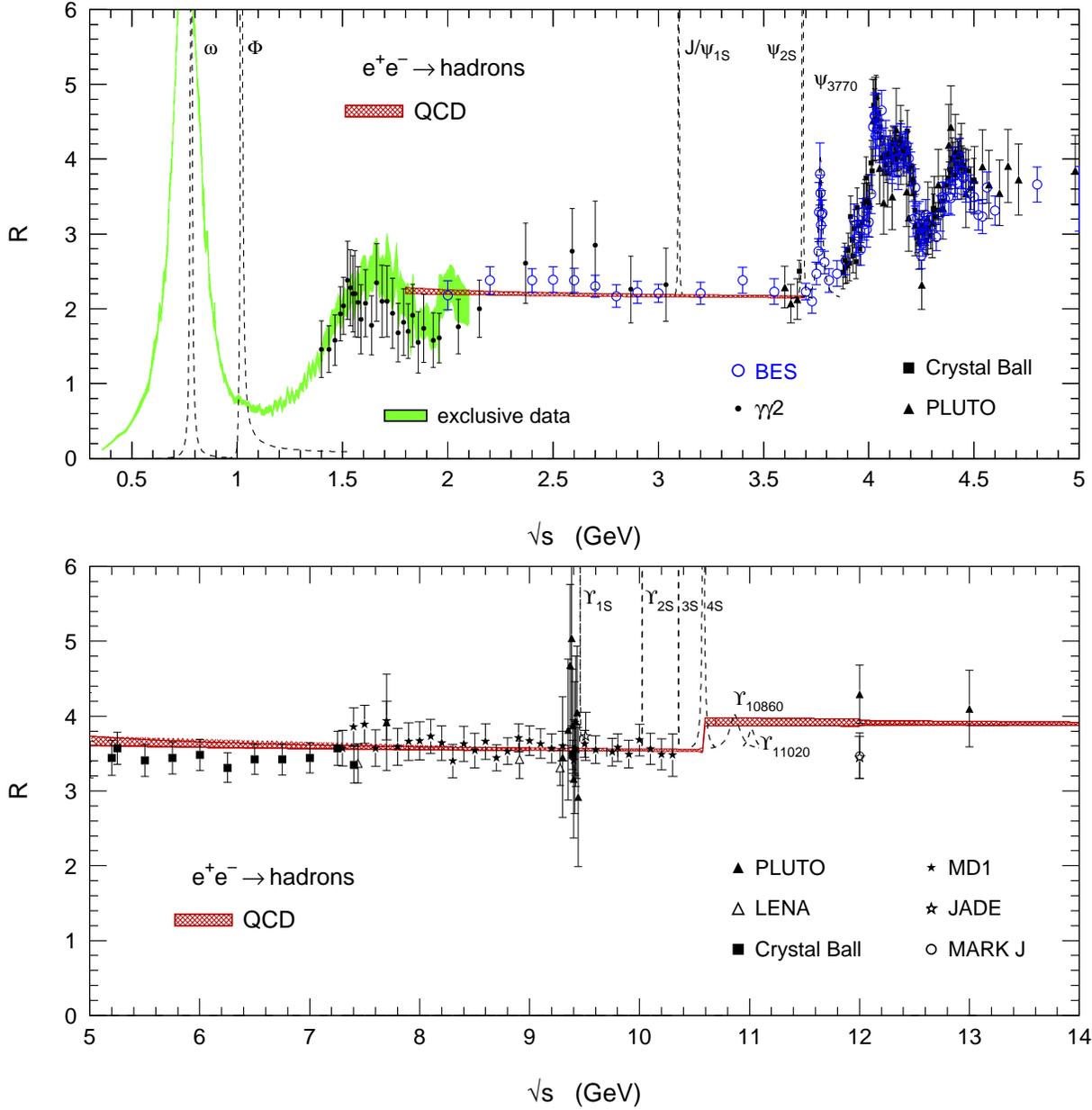}}
\caption[.]{\it Compilation of the data contributing to \amuhadLO. 
        Shown is the total hadronic over muonic cross section ratio $R$.
        The shaded band below 2~GeV represents the sum of the exclusive 
        channels considered in this analysis, with the exception of 
        the contributions from the narrow resonances which are given
        as dashed lines.
        All data points shown correspond to inclusive measurements. 
	The cross-hatched band gives the prediction
        from (essentially) perturbative QCD, which is found to be in good
        agreement with the measurements in the continuum 
	above 2~GeV. In this figure
        the $b\overline{b}$ threshold is indicated at the onset of 
        $B\overline{B}$ states in order to facilitate the comparison with 
        data in the
        continuum. In the actual calculation the threshold is taken at
        twice the pole mass of the $b$ quark.}
\label{fig_ree_all}
\end{figure}
\begin{table}[p]
\begin{center}
\setlength{\tabcolsep}{0.2pc}
{\small
\begin{tabular}{lcrrr} \hline 
&&&& \\[-0.2cm]
 & & \mc{3}{c}{\amuhadLO\ ($10^{-10}$)} \\
\rs{Modes} & \mc{1}{c}{\rs{Energy [GeV]}} & \mc{1}{c}{~\ee} 
	& \mc{1}{c}{$~\tau$\,$^(\footnotemark[3]{^)}$} 
		& \mc{1}{c}{$~\Delta(e^+e^--\tau)$} \\[0.15cm]
\hline
&&&& \\[-0.3cm]
Low $s$ exp. $\pi^+\pi^-$
	& $[2m_{\pi^\pm}-0.500]$   & $ 58.04\pm1.70\pm1.14$  
			& $ 56.03\pm1.61\pm0.28$ & $ +2.0\pm2.6$ \\
$\pi^+\pi^-      $  
	& $[0.500-1.800]$    & $440.81\pm4.65\pm1.54$  
			& $464.03\pm3.19\pm2.34$
                        & $-23.2\pm6.3$ \\
$\pi^0\gamma$, $\eta \gamma$\,$^(\footnotemark[1]{^)}$
	& $[0.500-1.800]$    & $  0.93\pm0.15\pm0.01$  & -  & - \\
$\omega$          
	& $[0.300-0.810]$    & $ 36.94\pm0.84\pm0.80$  & -  & - \\
$\pi^+\pi^-\pi^0$  {\footnotesize[below $\phi$]}
	& $[0.810-1.000]$    & $  4.20\pm0.40\pm0.05$  & -  & - \\
$\phi$  
	& $[1.000-1.055]$    & $ 34.80\pm0.92\pm0.64$  & -  & - \\
$\pi^+\pi^-\pi^0$  {\footnotesize[above $\phi$]}
	& $[1.055-1.800]$    & $  2.45\pm0.26\pm0.03$  & -  & - \\
$\pi^+\pi^-2\pi^0       $  
	& $[1.020-1.800]$    & $ 16.73\pm1.32\pm0.20$  
		& $ 21.44\pm1.33\pm0.60$
			& $ -4.7\pm1.8$ \\
$2\pi^+2\pi^-           $  
	& $[0.800-1.800]$    & $ 13.95\pm0.90\pm0.23$  
		& $ 12.34\pm0.96\pm0.40$
			& $ +1.6\pm2.0$ \\
$2\pi^+2\pi^-\pi^0        $  
	& $[1.019-1.800]$    & $  2.09\pm0.43\pm0.04$  & -  & - \\
$\pi^+\pi^-3\pi^0 $\,$^(\footnotemark[2]{^)}$  
	& $[1.019-1.800]$    & $  1.29\pm0.22\pm0.02$  & -  & - \\
$3\pi^+3\pi^-    $  
	& $[1.350-1.800]$    & $  0.10\pm0.10\pm0.00$  & -  & - \\
$2\pi^+2\pi^-2\pi^0       $  
	& $[1.350-1.800]$    & $  1.41\pm0.30\pm0.03$  & -  & - \\
$\pi^+\pi^-4\pi^0       $\,$^(\footnotemark[2]{^)}$    
	& $[1.350-1.800]$    & $  0.06\pm0.06\pm0.00$  & -  & - \\
$\eta${\footnotesize($ \rar\pi^+\pi^-\gamma$, $2\gamma$)}$\pi^+\pi^-$ 
	& $[1.075-1.800]$    & $  0.54\pm0.07\pm0.01$  & -  & - \\
$\omega${\footnotesize($\rar\pi^0\gamma$)}$\pi^{0}$
	& $[0.975-1.800]$    & $  0.63\pm0.10\pm0.01$  & -  & - \\
$\omega${\footnotesize($\rar\pi^0\gamma$)}$(\pi\pi)^{0}$
	& $[1.340-1.800]$    & $  0.08\pm0.01\pm0.00$  & -  & - \\
$K^+K^-            $  
	& $[1.055-1.800]$    & $  4.63\pm0.40\pm0.06$  & -  & - \\
$K^0_S K^0_L         $  
	& $[1.097-1.800]$    & $  0.94\pm0.10\pm0.01$  & -  & - \\
$K^0K^\pm\pi^\mp         $\,$^(\footnotemark[2]{^)}$    
	& $[1.340-1.800]$    & $  1.84\pm0.24\pm0.02$  & -  & - \\
$K\overline K\pi^0$\,$^(\footnotemark[2]{^)}$    
	& $[1.440-1.800]$    & $  0.60\pm0.20\pm0.01$  & -  & - \\
$K\overline K\pi\pi         $\,$^(\footnotemark[2]{^)}$    
	& $[1.441-1.800]$    & $  2.22\pm1.02\pm0.03$  & -  & - \\
$R=\sum{\rm excl.~modes}    $  
	& $[1.800-2.000]$    & $  8.20\pm0.66\pm0.10$  & -  & - \\
$R$ {\footnotesize[Data]}
	& $[2.000-3.700]$    & $ 26.70\pm1.70\pm0.00$  & -  & - \\
$J/\psi$         
	& $[3.088-3.106]$    & $  5.94\pm0.35\pm0.03$  & -  & - \\
$\psi(2S)$ 
	& $[3.658-3.714]$    & $  1.50\pm0.14\pm0.00$  & -  & - \\
$R$ {\footnotesize[Data]}  
	& $[3.700-5.000]$    & $  7.22\pm0.28\pm0.00$  & -  & - \\
$R_{udsc}$ {\footnotesize[QCD]}
	& $[5.000-9.300]$    & $  6.87\pm0.10\pm0.00$  & -  & - \\
$R_{udscb}$ {\footnotesize[QCD]}
	& $[9.300-12.00]$    & $  1.21\pm0.05\pm0.00$  & -  & - \\
$R_{udscbt}$  {\footnotesize[QCD]}
	& $[12.0-\infty]$    & $  1.80\pm0.01\pm0.00$  & -  & - \\[0.15cm]
\hline
&&&& \\[-0.3cm]
	&
			     & \mc{1}{r}{$684.7\pm6.0_{\rm exp}~~~$}  
			     & \mc{1}{l}{$709.0\pm5.1_{\rm exp}$} 
			     & \\
\rs{$\sum\;(e^+e^-\rightarrow\:$hadrons)}
 	& \rs{$[2m_{\pi^\pm}-\infty]$}
			     & \mc{1}{r}{$\pm\,3.6_{\rm rad\,}~~~$}  
			     & \mc{1}{r}{$\pm\,1.2_{\rm rad}\pm2.8_{\rm SU(2)}$} 
			     & \mc{1}{r}{\rs{$-24.3\pm7.9_{\rm tot}$}} 
	
\\[0.15cm]
 \hline
\end{tabular}
}
\end{center}
\vspace{-0.5cm}
{\footnotesize 
\begin{quote}
$^{1}\,$Not including $\omega$ and $\phi$ resonances (see text). \\ \noindent
$^{2}\,$Using isospin relations (see text). \\ \noindent
$^{3}\,$\ee\  data are used above 1.6~GeV (see text). \\ \noindent
\end{quote}
} 
\vspace{-0.75cm}
\caption{\label{tab_results}\em
	Summary of the \amuhadLO\ contributions from \ee\ 
        annihilation and $\tau$ decays. The uncertainties 
	on the vacuum polarization
	and FSR corrections are given as second errors in the individual 
        \ee\ contributions, while those from isospin breaking are 
        similarly given for the $\tau$ contributions. These 'theoretical'
        uncertainties are correlated among all channels, except in the
        case of isospin breaking which shows little correlation between 
        the $2\pi$ and $4\pi$ channels. The errors given 
        for the sums in the last line are from the experiment, the missing 
        radiative corrections in \ee\ and, in addition for $\tau$, SU(2)
        breaking.}
\end{table}
Fig.~\ref{fig_ree_all} gives a panoramic view of the \ee\ data in the 
relevant energy range. The shaded band below 2~GeV represents the sum 
of the exclusive channels considered in the analysis. It turns out to be
smaller than our previous estimate~\cite{adh}, 
essentially because more complete data sets are used and new information 
on the dynamics could be incorporated in the isospin
constraints for the missing channels. It should be pointed out that
the exclusive sum could lead to an underestimation of $R$, as
some unmeasured higher multiplicity hadronic channels 
could start to play a role in the 2~GeV region. Nevertheless, good
agreement is observed at 2~GeV with the first inclusive 
data point from BES, thus indicating that the missing component 
is likely to be small. The QCD prediction is indicated by the 
cross-hatched band. It is used in this analysis only for energies 
above 5~GeV. Note that the QCD band is plotted taking into account 
the thresholds for open flavour $B$ states, in order to facilitate 
the comparison with the data in the continuum. However, for the 
evaluation of the integral, the $b\overline{b}$ threshold is taken 
at twice the pole mass of the $b$ quark, so that the contribution 
includes the narrow $\Upsilon$ resonances, according to global 
quark-hadron duality.
\vs
The contributions from the different processes in their indicated 
energy ranges are listed in Table~\ref{tab_results}.
Wherever relevant, the two \ee- and $\tau$-based evaluations are given.
The discrepancies discussed above
are now expressed directly in terms of \amuhadLO\, giving smaller
estimates for \ee\ data by 
$(-21.2\pm6.4_{\rm exp}\pm2.4_{\rm rad}\pm2.6_{\rm SU(2)}\,(\pm7.3_{\rm total}))~10^{-10}$ for the $2\pi$ channel and 
$(-3.1\pm2.6_{\rm exp}\pm0.3_{\rm rad}\pm1.0_{\rm SU(2)}\,(\pm2.9_{\rm total}))~10^{-10}$ for the sum of the $4\pi$ channels. 
The total discrepancy  
$(-24.3\pm6.9_{\rm exp}\pm2.7_{\rm rad}\pm2.8_{\rm SU(2)}\,(\pm7.9_{\rm total}))~10^{-10}$ amounts to 3.1 standard deviations and
precludes from performing a straightforward combination of the two 
evaluations.

%
%
\subsection{\it Results for $a_\mu$}
\label{sec_results_amu}

\begin{figure}[t]
\epsfxsize12cm
\centerline{\epsffile{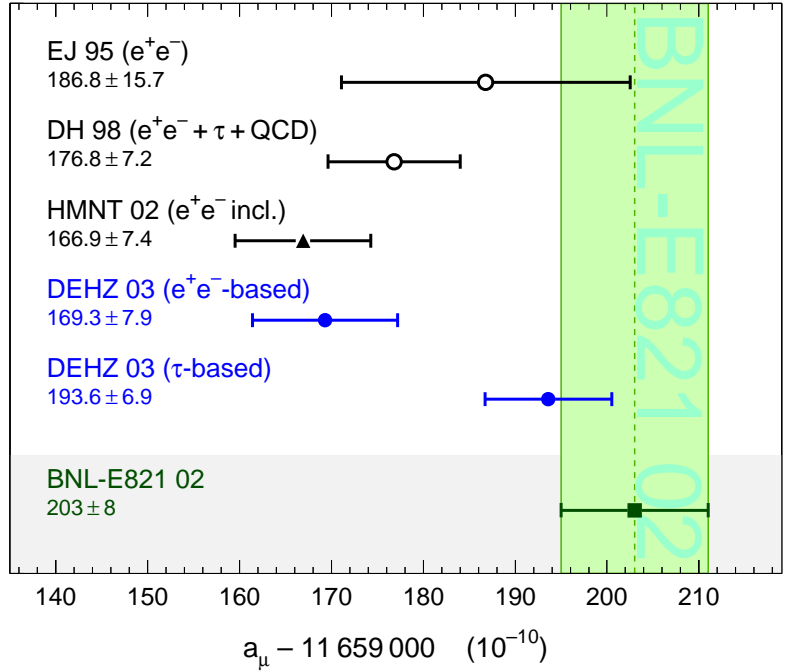}}
\caption[.]{\it Comparison of the results~(\ref{eq_smres}) with the 
	BNL measurement~\cite{bnl_2002}. Also shown
	are our previous estimates~\cite{eidelman,dh98} obtained before 
	the CMD-2 data were available, and the recent evaluation of 
	Hagiwara {\em et al.}~\cite{teubner}.}
\label{fig:results}
\end{figure}

The results for the lowest order hadronic contribution are \\[0.0cm]
\beq
 \begin{array}{|rcll|}
 \hline
  &&&\\[-0.1cm]
  ~~a_\mu^{\rm had,LO} &=& (684.7\pm6.0_{\rm exp}\pm3.6_{\rm rad})~10^{-10}
	&~~~[e^+e^-{\rm -based}]~~ \\[0.3cm]
 ~~a_\mu^{\rm had,LO} &=& (709.0\pm5.1_{\rm exp}\pm1.2_{\rm rad}
			\pm2.8_{\rm SU(2)})~10^{-10}
	&~~~[\tau{\rm -based}]~~ \\[0.3cm]
 \hline
  \mc{4}{c}{}\\[-0.1cm]
 \end{array}
\eeq
Adding the QED, higher-order hadronic, light-by-light scattering and
weak contributions as given in Section~\ref{anomaly},
the results for $a_\mu$ are obtained
\beqns
\label{eq_smres}
 \begin{array}{rcll}
  ~~a_\mu^{\rm SM} &=& (11\,659\,169.3\pm7.0_{\rm had}
		\pm3.5_{\rm LBL}\pm0.4_{\rm QED+EW})~10^{-10}
	&~~~[e^+e^-{\rm -based}]~,~ \\[0.2cm]
 ~~a_\mu^{\rm SM} &=& 	(11\,659\,193.6\pm5.9_{\rm had}
		\pm3.5_{\rm LBL}\pm0.4_{\rm QED+EW})~10^{-10}
	&~~~[\tau{\rm -based}]~.
 \end{array}
\eeqns
These values can be compared to the present experimental average given 
in Eq.~(\ref{bnl}). Adding experimental and theoretical errors 
in quadrature, the differences between measured and computed values 
are found to be:
\beq
 \begin{array}{rcll}
  a_\mu^{\rm exp}-a_\mu^{\rm SM} \,&=&\, (33.7\pm11.2)~10^{-10}
	&~~~[e^+e^-{\rm -based}]~, \\[0.2cm]
  a_\mu^{\rm exp}-a_\mu^{\rm SM} \,&=&\, (9.4\pm10.5)~10^{-10}
	&~~~[\tau{\rm -based}]~,
 \end{array}
\eeq
corresponding to 3.0 and 0.9 standard deviations, respectively.
A graphical comparison of the results~(\ref{eq_smres}) with the 
experimental value is given in Fig.~\ref{fig:results}. Also shown
are our previous estimates~\cite{eidelman,dh98} obtained before 
the CMD-2 and the new $\tau$ data were available (see discussion below), 
and the recent evaluation of Hagiwara {\em et al.}~\cite{teubner}.

\vfill
\pagebreak

%
%
\section{Discussion}
\label{sec_discuss}

%
%
\subsection{\it The Problem of the $2\pi$ Contribution}
\label{sec_discuss_2pi}

The significant discrepancy between the \ee\ and $\tau$ evaluations of 
\amuhadLO\ is a matter of concern. In this section we comment
on the relevant aspects of the problem. 
Since our earlier work~\cite{adh,dh97,dh98} was based on a combined 
analysis of \ee\ and $\tau$ data, we feel important to
summarize the main changes (all expressed in $10^{-10}$ units) in 
the dominant $2\pi$ contribution where
the $\tau$ contribution makes its impact:
\bei

\item the new CMD-2 data~\cite{cmd2} produce a downward shift of the 
\ee\ evaluation by 1.9 (well within errors from previous experiments), 
while the final error is reduced from $\pm12.5$ to $\pm5.1$ with an
additional $\pm2.4$ from missing radiative corrections, 

\item the new ALEPH data~\cite{aleph_new} increases the $\tau$ evaluation 
by 3.5, which is within the previous experimental uncertainty of $\pm$7.2
(in our previous analyses, we did not quote the results of a $\tau$-based
analysis alone, but only those from the combined \sfs),

\item including the CLEO data in the $\tau$ evaluation improves the 
precision, but further raises the central value by 4.0,

\item although including the OPAL data has little effect on the overall
precision, it also increases the result by 1.9,

\item the new complete isospin symmetry-breaking correction,
including the re-evaluation of the $S_{\rm EW}$ factor,
increases the $\tau$ evaluation by $0.2$ with respect to the 
previous one~\cite{adh,dh97,dh98}, which is well within the quoted 
error of $\pm2.5$.

\eei
The previous (unpublished) difference between the \ee- and $\tau$-based
evaluations of the $2\pi$ contributions, 
$\Delta \left(a_\mu^{\rm had,LO}\right)_{2\pi,ee-\tau}=
-10.8 \pm12.5_{\rm exp,ee}\pm7.2_{{\rm exp},\tau}\pm2.5_{{\rm SU(2)}}$, 
was consistent with zero, allowing the two \sfs\
to be combined into an improved common estimate. 

In spite of the fact that every change was within its 
previously estimated errors, the two results are not 
consistent anymore so that one must address the question 
of the possible origin of the problem.

In principle, the observed discrepancy for the $2\pi$ contribution,
$(-21.2\pm7.3)$, or $(-4.2\pm1.4)$\% when expressed 
with respect to \ee, could be caused by any (or the combination of 
several) of the following three effects which we examine in turn:
\bei

\item {\bf The normalization of \ee\  data}\\[0.05cm]
Here, as below, 'normalization' does not necessarily mean an overall
factor, but refers to the absolute scale of the 'bare' cross section
at each energy point. There is no cross check of this at the precision
of the new CMD-2 analysis. The only test we can provide is to compute
the \ee\ integral using the experiments separately. Because of the 
limited energy range where the major experiments overlap, we choose
to perform the integration in the range of $\sqrt{s}$ from 610.5 to 820~MeV.
The corresponding contributions are:
$313.5\pm3.1$ for CMD-2, $321.8\pm13.9$ for OLYA, $320.8\pm12.6$ for CMD,
and $323.9\pm2.1$ for the isospin-corrected $\tau$ data. 
No errors on radiative corrections and isospin breaking are
included in the above results.

\item {\bf The normalization of $\tau$ data}\\[0.05cm]
The situation is quite similar, as the evaluation is dominated by the
ALEPH data. It is also possible to compare the results provided by each
experiment separately, with the \sfs\ normalized to the respective
hadronic branching ratios. Leaving aside the region below 500~MeV where a fit
combining analyticity constraints is used, the contributions are:
$460.1\pm4.4$ for ALEPH, $464.7\pm9.3$ for CLEO and 
$464.2\pm8.1$ for OPAL, where the common error 
on isospin breaking has been left out.
The three values are consistent with each other and even the less 
precise values are not in good
agreement with the \ee\ estimate in this range, $440.8\pm4.7$,
not including the error on missing radiative corrections.
This is in line with the conclusion drawn from the comparison of
branching ratios presented in Fig.~\ref{fig_cvc_2pi}.
The larger values obtained with the CLEO and OPAL \sfs\  
are related to their relatively higher level below the $\rho$ 
resonance, as can be observed in Fig.~\ref{fig_2pi_tau}.

At the level of the $\tau\rightarrow\nu_\tau\pi^-\pi^0$ branching ratio,
which controls the normalization of the $\pi^-\pi^0$ \sf, stringent tests
can be applied to the ALEPH results. We stress the fact that the branching 
fractions are obtained by a global procedure where all $\tau$ decay final
states are considered, down to branching ratios of a few $10^{-4}$, from
a very clean initial sample~\cite{aleph_new}. The most critical part in 
the analysis is the separation of channels with different $\pi^0$
multiplicities. The $\pi^-\pi^0$ final state could be spoiled from the
adjacent channels $\pi^-$ and $\pi^-2\pi^0$ by inadequate
understanding of the $\gamma$ identification and the 
$\pi^0$ reconstruction. The observed branching
ratios for these two modes are in agreement with expectations, based for the
first one only on the assumption of universality of the $\mu-\tau$
couplings in the weak charged current (which is tested at the $3~10^{-3}$ 
level using the $\tau$ electronic branching ratio and the lifetime), and 
for the second one on the isospin relation with the $2\pi^-\pi^+$
branching ratio: 
$B_\pi-B_\pi^{\rm uni}=(-0.08\pm0.11_{\rm exp}\pm0.04_{\rm th})\%$,
and 
$B_{\pi 2\pi^0}-B_{\pi 2\pi^0}^{3\pi,{\rm iso}}=(+0.06\pm0.17_{\rm exp}\pm0.07_{\rm th})\%$.
These two tests provide confidence that the precise determination of the 
branching ratio for $\tau\rightarrow\nu_\tau\pi^-\pi^0$  
is on solid ground, as the observed discrepancy would require a 
shift of $1.1\%$ on this quantity.

Apart from an overall normalization effect, differences could originate
from the shape of the measured \sfs. If all three \sfs\ are normalized to
the world average branching ratio (our final procedure), then the results
for the contribution above 0.5~GeV become:
$459.9\pm3.6$ for ALEPH, $465.4\pm5.1$ for CLEO and 
$464.5\pm5.1$ for OPAL, with a common error of $\pm$2.4 from the $\pi \pi^0$ 
and leptonic branching ratios and the uncertainty on isospin breaking 
left out. Again the results are consistent and their respective experimental 
errors give a better feeling of the relative impact of the measurements.

\item {\bf The isospin-breaking correction applied 
	   to $\tau$ data}\\[0.05cm]
The basic components entering SU(2) breaking have been identified. The
weak points before were the poor knowledge of the long-distance 
radiative corrections and the quantitative effect of loops. 
Both points have been addressed by the analysis of
Ref.~\cite{ecker2} showing that the effects are small and covered by
the errors previously applied. The overall effect of the 
isospin-breaking corrections (including FSR)
applied to the $2\pi$ $\tau$ data, expressed in relative terms, is
$(-1.8\pm0.5)\%$. Its largest contribution ($-2.3\%$) 
stems from the uncontroversial
short-distance electroweak correction. Additional contributions 
must be identified to bridge the observed difference.

One could question the validity of the chiral model used. The authors of
Ref.~\cite{ecker2} argue that the corrections are insensitive to the 
details of their model and essentially depend only on the shape of the
pion form factor. As the latter is known from experiment to adequate
accuracy, there seems to be little space for improvement.

\eei

Thus we are unable at this point to identify the source of the discrepancy.
More experimental and theoretical work is needed. On the experimental side,
additional data is available from CMD-2, 
but not yet published. As an alternative, a promising approach 
using \ee\  annihilation events with initial state radiation (ISR), 
as proposed in Ref.~\cite{kuehn_isr}, allows a single experiment 
to cover simultaneously
a broad energy range. Two experimental programs are underway at Frascati
with KLOE~\cite{kloe_isr} and at SLAC with BABAR~\cite{babar_isr}. The
expected statistics are abundant, but it will be a challenge to reduce
the systematic uncertainty at the level necessary to probe the CMD-2
results. However, the
experimental technique being so different, it will be in any case
valuable to compare the results with the present ones. As for $\tau$'s,
the attention is now focused on the forthcoming results from the $B$ 
factories. Again, the quality of the analysis will be determined 
by the capability to control systematics rather than the already sufficient
statistical accuracy.
On the theory side, the computation of more precise and more
complete radiative corrections both for \ee\ cross sections and 
$\tau$ decays should be actively pursued.  

%
%
\subsection{\it Other Points of Discussion}
\label{sec_discuss_other}

Other points are worth to be discussed: the $4\pi$ \sfs, the $\omega$
and $\phi$ resonances and the sum of exclusive channels from 1.6 to 2 GeV.
\vs
As already pointed out in Sections~\ref{sec_compsf} and \ref{sec_brcvc},
the quality of both \ee\ and $\tau$ data in the $2\pi^+2\pi^-$ and
$\pi^+\pi^-2\pi^0$ final states is not as good as for the 
$\pi^+\pi^-$ channel. Agreement is observed in the former channel at the 
$10\%$ level, while in the latter a large discrepancy is found (see the
values in Table~\ref{tab_results}) which 
to this level cannot be attributed to isospin breaking.
Since significant differences are found within the \ee\ data sets, we
feel that it is a priority to clarify the experimental situation 
in this sector. 
The ISR program being conducted with the BABAR experiment should be able 
to shed some light upon this problem~\cite{babar_isr}. 
\vs
Compared to previous estimates, the $\omega$ and $\phi$ resonance
contribution is now directly evaluated with the measured cross section, 
rather than integrating a Breit-Wigner function computed with averaged
parameters. The $\omega$ value is basically unchanged, while a large 
downward shift of 4.3 has been found for the $\phi$ contribution. The 
origin of this change lies in the fact that the recent measurement
of the $\phi$ lineshape yields a total width which is significantly
smaller: $\Gamma_\phi$ decreased by $6\sigma$ in the last two 
years~\cite{pdg2002}! 
\vs
Revisiting the situation of the exclusive channels in the 1.6-2~GeV range
has led to significant changes, the origin of which are twofold:
({\it i}) more information (obtained in the study of $\tau$ decays, see 
Section~\ref{sec_dat_ee}) on the decay dynamics in the $6\pi$ channel could
be used to bound the $\pi^+\pi^-4\pi^0$ contribution, and ({\it ii}) some
data with poor quality were discarded, resulting in smaller contributions
in the $3\pi^+3\pi^-$ and $2\pi^+2\pi^-2\pi^0$
channels. As a result, and unlike the conclusion reached
in Refs.~\cite{martin_alpha,teubner}, we find the sum of the 
exclusive processes to be in reasonable agreement with the 
inclusive measurements of $R$ in this range~\cite{E_78}. 
At any rate it is clear that better data should be taken in this 
energy region. The BABAR ISR physics program should be able to 
make an important contribution here as well.
\vs
Due to the last two points which are only relevant to \ee\ 
data---the contribution from the $\phi$ resonance and the 
multi-pion channels---our new evaluation comes out to be 
significantly smaller than before. 

%
%
\subsection{\it Comparison to Other Evaluations of \amuhadLO}
\label{sec_discuss_compet}

Here we restrict our discussion to recent evaluations which have been
published since 1998, {\it i.e.} Refs.~\cite{narison,troconiz}. 
Previous estimates were considered in our earlier publication~\cite{dh98}.
A common feature of Refs.~\cite{narison,troconiz} is that they use both
\ee\ and $\tau$ data for the $2\pi$ contribution. However, their analyses 
are based on the preliminary CMD-2 data~\cite{cmd2_prel} which are 
not corrected for vacuum polarization and FSR. Because of this, they fail to
notice the discrepancy between \ee\ and $\tau$ data. In addition, no mention
of isospin symmetry breaking, and how to correct for it, is made in 
Ref.~\cite{narison}, shedding some doubts about the validity of the
combination of the two data sets. The relatively high values obtained in
these analyses compared to the present one are due in part to these
problems.

The recent analysis of Hagiwara {\em et al.}~\cite{teubner} does include 
the final CMD-2 data. Our \ee-based result agrees with their
evaluation using inclusive hadron production for energies above 1.6~GeV. 
However, as pointed out before, our re-evaluated sum of exclusive channels 
in this range is consistent within errors with the inclusive rate.

%
%
\subsection{\it Consequences for \aqedZ}
\label{sec_results_alpha}

In spite of the fact that the present analysis was focused on the
theoretical prediction for the muon magnetic anomaly, it is possible
to draw some conclusions relevant to the evaluation of the hadronic vacuum
polarization correction to the fine structure constant at $M_Z^2$. 
The problem found in the $2\pi$ \sf\  is less important for 
\daqedZ\  with respect to the total uncertainty,
because the integral involved gives less weight to the low-energy region. 
The difference between the evaluations using the $2\pi$, 
$4\pi$ and $2\pi2\pi^0$ \sfs\ from \ee\ and $\tau$ data are found to be:
\beqns
 \Delta\alpha^{ee}_{\rm had}(M_{Z}^2)\,-\,
 \Delta\alpha^{\tau}_{\rm had}(M_{Z}^2)= (-2.79\pm
    \underbrace{0.43_{\rm ee}\pm0.26_{\rm rad}
	\pm0.55_\tau\pm0.30_{\rm SU(2)}}_{0.80})~10^{-4}~.
\eeqns
While this low-energy contribution shows a 3.5 standard deviation 
discrepancy (when adding the different errors in quadrature), it also 
exceeds the total uncertainty of $1.6~10^{-4}$ 
on \daqedZ\ which was quoted in Ref.~\cite{dh98}. It is worth pointing out
that such a shift produces a noticeable effect for the determination of the 
Higgs boson mass $M_{\rm H}$ in the global electroweak fit~\cite{haidt}. 
With the present input for the electroweak observables~\cite{lepewwg} 
from LEP, SLC and FNAL yielding central values for $M_{\rm H}$ around 100~GeV,
going from the \ee\- to the $\tau$-based evaluation induces
a decrease of $M_{\rm H}$ by 16~GeV using all observables and by 20~GeV when
only the most sensitive observable, $(\sin^2\theta_{\rm W})_{\rm eff}$, 
is used.

%
%
\section{Conclusions}

A new analysis of the lowest-order hadronic vacuum polarization contribution
to the muon anomalous magnetic moment has been performed. It is based on the
most recent high-precision experimental data from \ee\ annihilation and 
$\tau$ decays in the $\pi\pi$ channel. Special attention was given to 
the problem of isospin symmetry breaking and the corresponding 
corrections to be applied to $\tau$ data. A new theoretical analysis 
of radiative corrections in $\tau$ decays was used and found to be 
in agreement with previous estimates. A complete re-evaluation of the
contributions of \ee\ annihilation cross sections in the energy range
up to 2~GeV has been performed. Incorporating the recently corrected
contribution from light-by-light scattering diagrams, the full prediction
for the muon magnetic anomaly $a_\mu$ is obtained.
\vs
The main results of our analysis are the following:

\bei
\item the new evaluation based solely on \ee\ data is significantly
lower than previous estimates and is in conflict with the experimental
determination of $a_\mu$ by 3.0 standard deviations.
\item the new precise evaluations of the dominant $\pi\pi$ contributions 
from \ee\ annihilation and isospin-breaking corrected $\tau$ decays are 
not anymore in agreement with each other. A discussion has been presented 
for possible sources of the discrepancy which could not be resolved.
This situation is a matter of great concern, as the $\tau$-based prediction
of $a_\mu$ is in better agreement with the experimental value, 
from which it deviates by non-significant 0.9 standard deviations.
\eei
More experimental and theoretical work is needed to lift the present
uncertainty on whether or not new physics has been uncovered with the
muon magnetic moment.  
%
%
\subsection*{Acknowledgements}

It is a pleasure to thank our experimental colleagues from INP Novosibirsk,
IHEP Beijing, LEP and Cornell for their essential contributions, and 
members of the Muon $(g-2)$ Collaboration for their keen interest. 
The close cooperation with V.~Cirigliano, G.~Ecker and H.~Neufeld 
is warmly acknowledged. Informative discussions with F.~Jegerlehner, 
J.~H.~K\"uhn, A.~Pich, A.~Stahl, A.~Vainshtein and especially 
W.~Marciano are appreciated. We thank S.~Menke for providing us 
with the OPAL spectral functions.
\vfill
\pagebreak
%
%
\end{document}